\documentclass[twocolumn]{aastex631}

\newcommand\aastex{AAS\TeX}

\accepted{November 4, 2025}
\submitjournal{AAS}

\shorttitle{\aastex\ ALMA sub-pc resolution dense molecular line 
observations of NGC 1068}  
\shortauthors{Imanishi et al.}

\begin{document}

\title{ALMA Sub-parsec Resolution Dense Molecular Line Observations of
the NGC 1068 Nucleus}

\correspondingauthor{Masatoshi Imanishi}
\email{masa.imanishi@nao.ac.jp, m.imanishi.astro@gmail.com}

\author[0000-0001-6186-8792]{Masatoshi Imanishi}
\affil{National Astronomical Observatory of Japan, National Institutes 
of Natural Sciences (NINS), 2-21-1 Osawa, Mitaka, Tokyo 181-8588, Japan}
\affil{Department of Astronomy, School of Science, Graduate
University for Advanced Studies (SOKENDAI), Mitaka, Tokyo 181-8588,
Japan} 
\affil{Toyo University, 5-28-20, Hakusan, Bunkyo-ku, Tokyo 112-8606, 
Japan}

\author{Bernd Vollmer}
\affil{Universite de Strasbourg, CNRS, Observatoire astronomique de
Strasbourg, UMR 7550, 67000, Strasbourg, France} 

\author[0000-0002-9043-6048 ]{Yoshiaki Hagiwara}
\affil{Natural Science Laboratory, Toyo University, 5-28-20, Hakusan, 
Bunkyo-ku, Tokyo 112-8606, Japan} 

\author[0000-0002-6939-0372]{Kouichiro Nakanishi}
\affil{National Astronomical Observatory of Japan, National Institutes 
of Natural Sciences (NINS), 2-21-1 Osawa, Mitaka, Tokyo 181-8588, Japan}
\affil{Department of Astronomy, School of Science, Graduate
University for Advanced Studies (SOKENDAI), Mitaka, Tokyo 181-8588,
Japan} 

\author[0000-0001-9452-0813]{Takuma Izumi}
\affil{National Astronomical Observatory of Japan, National Institutes 
of Natural Sciences (NINS), 2-21-1 Osawa, Mitaka, Tokyo 181-8588, Japan}
\affil{Department of Astronomy, School of Science, Graduate
University for Advanced Studies (SOKENDAI), Mitaka, Tokyo 181-8588,
Japan} 

\author[0000-0003-2535-5513]{Nozomu Kawakatu}
\affil{Faculty of Natural Sciences, National Institute of Technology, 
Kure College, 2-2-11 Agaminami, Kure, Hiroshima 737-8506, Japan}



\begin{abstract}

We present the results of our ALMA observations of the dense molecular 
HCN J=4--3 and HCO$^{+}$ J=4--3 lines at $\lesssim$1 pc 
($\lesssim$14 mas) resolution in the nuclear region of the nearby 
($\sim$14 Mpc) well-studied AGN NGC 1068. 
Both emission lines are clearly detected around the AGN along 
an almost east-west direction, which we ascribe to the dusty 
molecular torus. 
The HCN J=4--3 emission is brighter than the HCO$^{+}$ J=4--3 emission 
in the compact ($\lesssim$3--5 pc) torus region. 
Apparent counter-rotation between the inner ($\lesssim$2 pc) and 
outer ($\gtrsim$2 pc) parts of the western torus, previously seen 
in $\sim$1.5 pc-resolution HCN J=3--2 and HCO$^{+}$ J=3--2 data, 
is also confirmed in our new $\lesssim$1 pc-resolution 
HCN J=4--3 and HCO$^{+}$ J=4--3 data. 
We apply a physically counter-rotating torus model, in which 
a compact dense gas clump collided with the western side of the 
existing rotating torus from the opposite direction, and 
we find that this model largely reproduces the observed
properties of the combined new $\lesssim$1 pc-resolution HCN J=4--3
and HCO$^{+}$ J=4--3 data, and the previously obtained 
$\lesssim$1.5 pc-resolution HCN J=3--2 and HCO$^{+}$ J=3--2 data.

\end{abstract}




\section{Introduction} 

An active galactic nucleus (AGN) shines brightly in the
UV--optical because of strong emission from an accretion disk around a
mass-accreting supermassive black hole (SMBH).
Some AGNs show broad ($\gtrsim$1000 km s$^{-1}$) emission
lines in the UV--optical spectra (classified as type 1), while others
do not (type 2).
These AGN properties can naturally be explained by the presence of
toroidally distributed, optically thick dust and gas on a $\lesssim$10
pc physical scale, the so-called dusty molecular torus 
\citep[e.g.,][]{ant93,hon19}.
Fast-rotating gas clouds in the vicinity of the SMBH, dominated by its
gravity and photoionized by UV radiation from the accretion disk,
can display broad ($\gtrsim$1000 km s$^{-1}$) UV--optical emission lines.
They are detectable if seen from a direction not
blocked by the torus (type 1) but become undetectable if obscured by
the torus along our line of sight (type 2).
This torus-based AGN unification picture was first proposed based on
observations of a very nearby ($\sim$14 Mpc, $z =$ 0.0038,
1 arcsec = 70 pc) type 2 AGN, NGC 1068 \citep{ant85}.
Detailed observations of the torus in NGC 1068 are thus very important
to better understand the torus-based AGN unification paradigm and
update it if necessary.

A spatially resolved observational study of the torus is obviously one
of the best means to investigate its properties in detail.
For this purpose, high-angular-resolution observations with
$\lesssim$0$\farcs$05 are desirable, because the physical size of the
putative torus is $\lesssim$10 pc, which corresponds to
$\lesssim$0$\farcs$15 even for very nearby ($\sim$15 Mpc) AGNs.
ALMA high-angular-resolution ($\lesssim$0$\farcs$05) molecular line
observations are a very powerful tool to scrutinize the morphological,
physical, chemical, and dynamical properties of the torus, because (1)
very high angular resolution ($\lesssim$0$\farcs$02) and high
sensitivity are achievable and (2) multiple rotational (J) transition
line observations of multiple molecules enable us to investigate the
density, temperature, abundance, and kinematics of molecular gas in
the torus.
Previously conducted ALMA high-angular-resolution
($\sim$0$\farcs$02--0$\farcs$04 [20--40 mas] or 1.5--3 pc) molecular
line observations of NGC 1068, mainly using the J=3--2 lines of CO,
HCN, and HCO$^{+}$, have revealed the presence of compact
($\lesssim$3--5 pc) dense and warm molecular gas along an almost
east-west direction
\citep[e.g.,][]{gar16,ima18a,gar19,imp19,ima20}, which can naturally
be interpreted as originating in the putative dusty molecular torus.
However, the observed torus molecular gas properties of NGC 1068 are
not in perfect agreement with the simple theory that predicts that
molecular gas in the torus should rotate with an almost Keplerian motion
dominated by the gravity of the central SMBH \citep[e.g.,][]{wad09,wad16}.
It was found that the inner ($\lesssim$2 pc) and outer ($\gtrsim$2 pc)
parts of the NGC 1068 torus do not rotate in the same way, but appear
to be counter-rotating relative to each other 
\citep[e.g.,][]{gar19,imp19,ima20}.
A first proposed scenario to explain this peculiar observational
result is that a collision of a very compact gas clump with the
western side of the existing rotating torus from the opposite
direction could create a bona-fide counter-rotating molecular torus
\citep[e.g.,][]{imp19,ima20,vol22}.
A second scenario is that outflow activity could reproduce the
observed apparent counter-rotation patterns 
\citep[e.g.,][]{gar19,wil20,ban23}.

The achievable angular resolution can be even smaller for the J=4--3
lines of HCN and HCO$^{+}$ at $\sim$0.85 mm ($\sim$350 GHz) than their
J=3--2 lines at $\sim$1.2 mm ($\sim$260 GHz), because of the shorter
wavelength (higher frequency) of the J=4--3 transitions.
Using the longest-baseline ($\sim$16 km) observations of ALMA,
we can achieve $\lesssim$0$\farcs$014 ($\lesssim$14 mas) angular
resolution for the J=4--3 lines, which corresponds to sub-parsec
($\lesssim$1 pc) physical resolution for NGC 1068 (1 arcsec = 70 pc).
Very recently, \citet{gam25} presented $\sim$0$\farcs$016
($\sim$1.1 pc) resolution HCO$^{+}$ J=4--3 and CO J=3--2 line data of 
NGC 1068, and confirmed the apparent counter-rotation between the inner
($\lesssim$2 pc) and outer ($\gtrsim$2 pc) torus regions,
particularly clearly in the western torus.
These authors proposed a new third scenario in which an infalling
compact gas cloud passes from the west, in front of the $\sim$0.85 mm
($\sim$350 GHz) continuum emitting central engine of the AGN (=
mass-accreting SMBH), without strongly interacting with the dusty
molecular torus.
Because (1) HCN emission is brighter than HCO$^{+}$ emission in
the compact ($\lesssim$3--5 pc) molecular torus of NGC 1068, 
most likely because of HCN abundance enhancement
\citep{ima18a,ima20,vol22,but22} and (2) HCN and HCO$^{+}$ have different
critical densities \citep{shi15}, the highest-angular-resolution 
HCN J=4--3 line data will enable us to better elucidate the dense
molecular gas properties of the NGC 1068 torus.
The addition of high-angular-resolution J=4--3 data of HCN and
HCO$^{+}$ to the existing $\sim$1.5 pc resolution J=3--2 data of HCN
and HCO$^{+}$ \citep[e.g.,][]{imp19,ima20} can also provide important
information on molecular gas excitation conditions inside the
$\sim$3--5 pc extended NGC 1068 torus in a spatially resolved manner.
In this paper, we present ALMA $\lesssim$1 pc resolution HCN J=4--3
and HCO$^{+}$ J=4--3 line observations of the NGC 1068 nucleus to
further scrutinize various properties of the dense molecular torus around
the AGN in NGC 1068.

\section{Observations and Data Analysis} 

Our HCN J=4--3 and HCO$^{+}$ J=4--3 line observations of the NGC 1068
nucleus were conducted in our ALMA Cycle 9 program 2022.1.00005.S 
(PI = M. Imanishi).  
To achieve the highest angular resolution, the longest baseline
($\sim$16 km) configuration was used.
We employed the widest 1.875 GHz mode with 1920 channels for each
spectral window. 
The HCN J=4--3 (rest frame frequency $\nu_{\rm rest}$ = 354.505 GHz)
and HCO$^{+}$ J=4--3 ($\nu_{\rm rest}$ = 356.734 GHz) lines were
simultaneously observed in one sideband (USB). 
The vibrationally excited HCN v$_{2}$=1, l=1f (HCN-VIB) J=4--3
line ($\nu_{\rm rest}$ = 356.256 GHz) is also covered in the USB,
while the vibrationally excited HCO$^{+}$ v$_{2}$=1, l=1f 
(HCO$^{+}$-VIB) J=4--3 line ($\nu_{\rm rest}$ = 358.242 GHz) is not covered. 
In LSB, the CS J=7--6 line ($\nu_{\rm rest}$ = 342.883 GHz) is
covered.
Table \ref{tab:obs} summarizes our ALMA Cycle 9 observation log. 

\begin{deluxetable*}{llccc|ccc}[!bht]
\tabletypesize{\scriptsize}
\tablecaption{Log of ALMA Cycle 9 Observations of NGC 1068 \label{tab:obs}} 
\tablewidth{0pt}
\tablehead{
\colhead{Line} & \colhead{Date} & \colhead{Antenna} & 
\colhead{Baseline} & \colhead{Integration} & \multicolumn{3}{c}{Calibrator} \\ 
\colhead{} & \colhead{(UT)} & \colhead{Number} & \colhead{(m)} &
\colhead{(min)} & \colhead{Bandpass} & \colhead{Flux} & \colhead{Phase}  \\
\colhead{(1)} & \colhead{(2)} & \colhead{(3)} & \colhead{(4)} &
\colhead{(5)} & \colhead{(6)} & \colhead{(7)}  & \colhead{(8)} 
}
\startdata 
HCN/HCO$^{+}$ J=4--3 & 2023 July 28 & 48 & 230--16196 & 44 & 
J0238$+$1636 & J0238$+$1636 & J0239$-$0234 \\  
 & 2023 July 29 & 44 & 230--16196 & 44 
& J0238$+$1636 & J0238$+$1636 & J0239$-$0234 \\  
 & 2023 July 31 & 46 & 230--16196 & 44 & J0238$+$1636 & J0238$+$1636 &
J0239$-$0234 \\  
 & 2023 August 3 & 49 & 230--16196 & 44 & J0238$+$1636 & J0238$+$1636 &
J0239$-$0234 \\  
 & 2023 August 4 & 49 & 230--16196 & 44 & J0238$+$1636 & J0238$+$1636 &
J0239$-$0234 \\  
 & 2023 August 9 & 44 & 83--16196 & 44 & J0238$+$1636 & J0238$+$1636 &
J0239$-$0234 \\  
 & 2023 August 11 & 41 & 83--16196 & 44 & J0238$+$1636 & J0238$+$1636 &
J0239$-$0234 \\  
\enddata

\tablecomments{
Col.(1): Line. 
Col.(2): Observation date in UT. 
Col.(3): Number of antennas used for observations. 
Col.(4): Baseline length in meters. Minimum and maximum baseline lengths are 
shown.  
Col.(5): Net on source integration time in minutes.
Cols.(6), (7), and (8): Bandpass, flux, and phase calibrators for the 
target source (NGC 1068), respectively.
}

\end{deluxetable*}


We started our analysis with pipeline-calibrated data provided by
ALMA, using CASA 6.5.4-9 \citep{CASA22,hun23}.
We determined the continuum level by selecting channels that did not
show obvious emission or absorption lines and subtracted it using the
CASA task ``uvcontsub''.
Then we applied the ``tclean'' task (Briggs-weighting; robust $=$ 0.5,
gain $=$ 0.1) to the continuum-subtracted molecular line and continuum
data, with a pixel scale of 0$\farcs$0015 pixel$^{-1}$ (1.5 mas
pixel$^{-1}$) and a velocity resolution of 10 km s$^{-1}$.
The derived rms noise level for the 10 km s$^{-1}$ velocity
resolution is $\sim$0.3 mJy beam$^{-1}$ for both HCN J=4--3 and
HCO$^{+}$ J=4--3.
The maximum recoverable scale (MRS) of our ALMA Cycle 9 data is
$\gtrsim$0$\farcs$14 (ALMA Cycle 9 Proposer's Guide), which corresponds
to $\gtrsim$10 pc at the distance of NGC 1068 and is thus large
enough to properly investigate the morphology and kinematics of the
compact ($\lesssim$3--5 pc) dense molecular torus around the AGN in
NGC 1068 \citep[e.g.,][]{ima16a,gar16,ima18a,gar19,imp19,ima20,vol22,gam25}.
The absolute flux calibration uncertainty is expected to be
$\lesssim$10\% at the observed frequency ($\sim$350 GHz) of the ALMA
Cycle 9 data.

\citet{ima20} presented cleaned maps of $\sim$1.3 pc
($\sim$0$\farcs$019) resolution HCN J=3--2 ($\nu_{\rm rest}$ = 265.886
GHz) and HCO$^{+}$ J=3--2 ($\nu_{\rm rest}$ = 267.558 GHz) line data
taken in ALMA Cycle 6 (2018.1.00037.S), with pixel scale of
0$\farcs$003 pixel$^{-1}$ (3 mas pixel$^{-1}$).
We reanalyzed the HCN J=3--2 and HCO$^{+}$ J=3--2 data with pixel
scale of 1.5 mas pixel$^{-1}$ and velocity resolution of 10 km
s$^{-1}$ to compare them with the newly taken, $\lesssim$1 pc
resolution HCN J=4--3 and HCO$^{+}$ J=4--3 line data in a
straightforward manner.
The derived rms noise level for the 10 km s$^{-1}$ velocity
resolution is $\sim$0.2 mJy beam$^{-1}$ for both HCN J=3--2 and
HCO$^{+}$ J=3--2.
The absolute flux calibration uncertainty of the ALMA Cycle 6 data at
$\sim$260 GHz is also expected to be $\lesssim$10\% (ALMA Cycle 6
Proposer's Guide).

\section{Results}

\begin{figure*}[!hbt]
\begin{center}
\hspace{-0.4cm}
\includegraphics[scale=0.37]{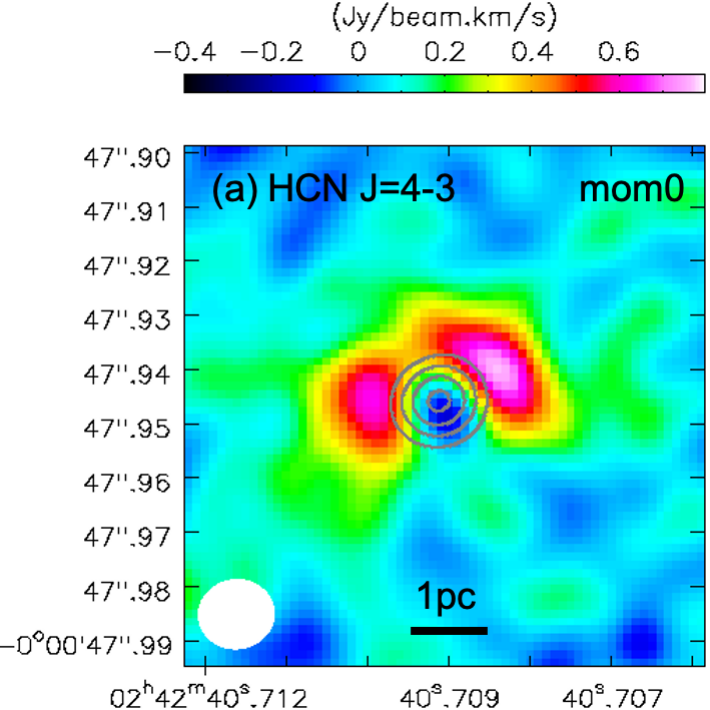} 
\includegraphics[scale=0.37]{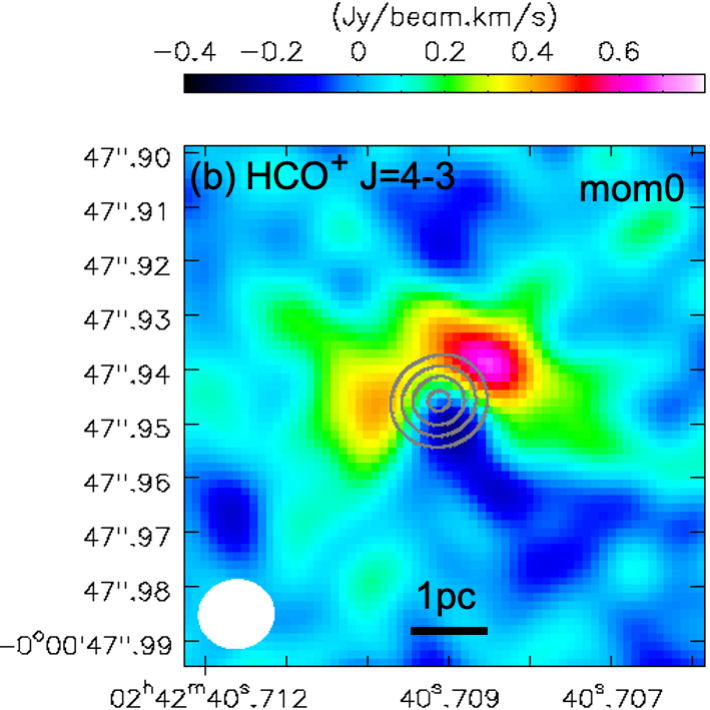} 
\includegraphics[scale=0.37]{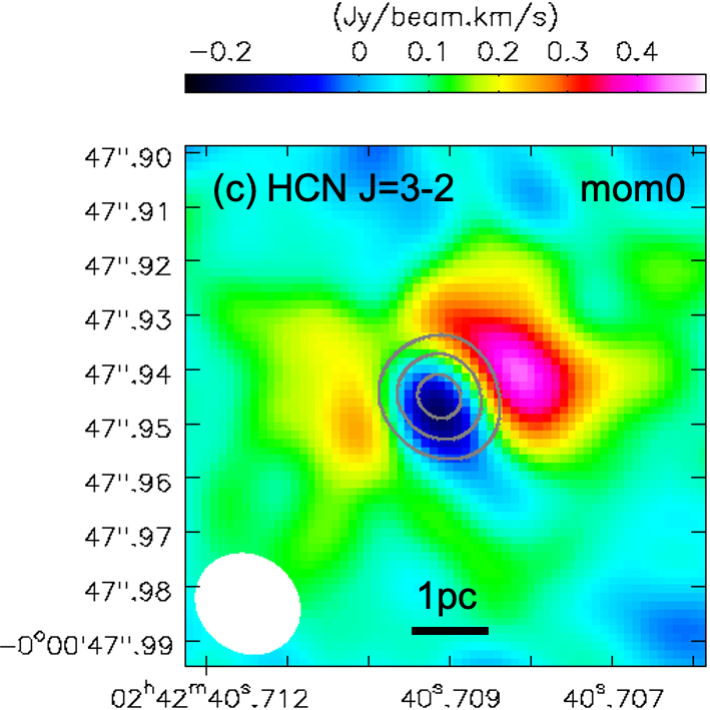} 
\includegraphics[scale=0.37]{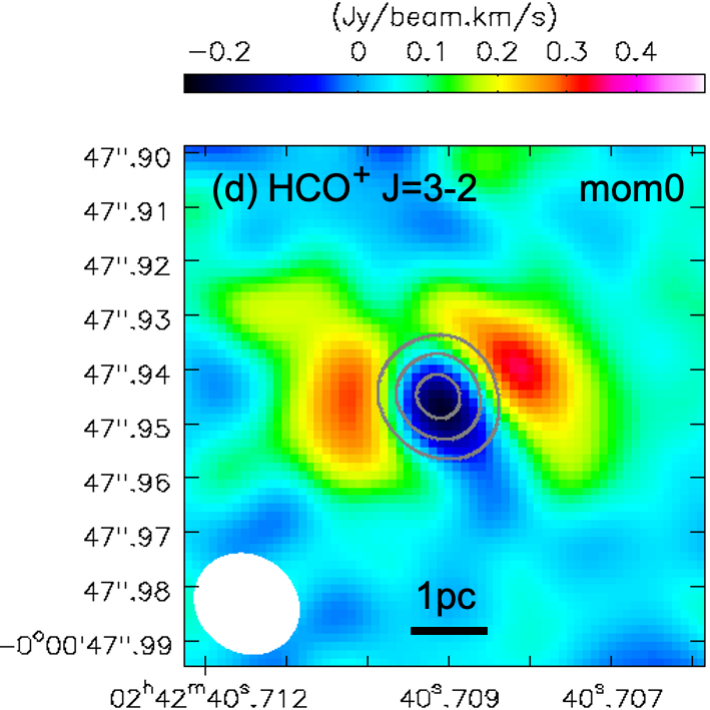} 
\end{center}
\vspace{-0.2cm}
\caption{
Integrated-intensity (moment 0) map of (a) HCN J=4--3,
(b) HCO$^{+}$ J=4--3, (c) HCN J=3--2, and (d) HCO$^{+}$ J=3--2,
created with a pixel scale of 1.5 mas pixel$^{-1}$.
The gray contours indicate simultaneously obtained continuum emission with
50$\sigma$, 70$\sigma$, 90$\sigma$, and 110$\sigma$ (a--b at $\sim$350 GHz)
and 40$\sigma$, 60$\sigma$, and 80$\sigma$ (c--d at $\sim$260 GHz), 
where the 1$\sigma$ value is shown in Table \ref{tab:cont}.
The horizontal thick black bar at the bottom center of each panel
corresponds to 1 pc at the distance of NGC 1068.
The filled white circle at the lower left of each panel indicates
the synthesized beam size, which is 0$\farcs$014 $\times$ 0$\farcs$013
for the J=4--3 lines of HCN and HCO$^{+}$ (a--b), and 0$\farcs$020 $\times$
0$\farcs$018 for the reanalyzed images of the J=3--2 lines (c--d).
\label{fig:mom0}
}
\end{figure*}

Table \ref{tab:cont} summarizes the continuum emission properties at
$\sim$350 GHz and the synthesized beam size (0$\farcs$014 $\times$
0$\farcs$013) as well as the reanalyzed $\sim$260 GHz data.
Figure \ref{fig:mom0}a and \ref{fig:mom0}b show $\sim$350 GHz
continuum (contours) and integrated-intensity (moment 0) maps of HCN
J=4--3 and HCO$^{+}$ J=4--3 lines, respectively.
Almost east-west oriented, compact ($\lesssim$3--5 pc) HCN J=4--3 and
HCO$^{+}$ J=4--3 emission is clearly detected around the continuum
peak (defined as ``C-peak''), which we regard as the position of the
mass-accreting SMBH.
We find that (1) the HCN J=4--3 emission is brighter than the HCO$^{+}$
J=4--3 emission in both the western and eastern torus, and
(2) the western torus is brighter than the eastern torus for both the HCN
J=4--3 and HCO$^{+}$ J=4--3 lines.
Continuum emission at $\sim$260 GHz (contours) and reanalyzed moment 0
maps of HCN J=3--2 and HCO$^{+}$ J=3--2 lines are displayed in
Figure \ref{fig:mom0}c and \ref{fig:mom0}d, respectively.
In Figure \ref{fig:mom0}a--d, the J=4--3 emission is spatially more
compact than the J=3--2 emission for both HCN and HCO$^{+}$.
Although the smaller synthesized beam size at J=4--3 could 
contribute to the observed more compact emission, we interpret that 
the J=4--3 lines probe warmer and denser molecular gas closer to the
central AGN engine than the J=3--2 lines because of the higher
excitation energy and critical density of the former (see the last
paragraph of $\S$3).

\begin{deluxetable*}{ccccc}[!hbt]
\tabletypesize{\scriptsize}
\tablecaption{Continuum Emission \label{tab:cont}} 
\tablewidth{0pt}
\tablehead{
\colhead{Frequency} & \colhead{Peak flux} & 
\colhead{Peak coordinate} & \colhead{rms} & \colhead{Synthesized beam} \\
\colhead{(GHz)} & \colhead{(mJy/beam)} & 
\colhead{(RA,DEC)ICRS} & \colhead{(mJy/beam)} & 
\colhead{($''$ $\times$ $''$) ($^{\circ}$)} \\  
\colhead{(1)} & \colhead{(2)} & \colhead{(3)}  & \colhead{(4)}  &
\colhead{(5)} 
}
\startdata 
340.4--344.2, 352.4--356.2 & 6.8 (116$\sigma$) & 
(02$^{\rm h}$ 42$^{\rm m}$ 40.709$^{\rm s}$, $-$00$^{\circ}$ 00$'$
47$\farcs$946) (C-peak) & 0.059  
& 0.014$\times$0.013 ($-$86$^{\circ}$) \\
263.6--269.1 & 6.7 (89$\sigma$) & (02$^{\rm h}$ 42$^{\rm m}$
40.709$^{\rm s}$, $-$00$^{\circ}$ 00$'$ 47$\farcs$946) & 0.075 & 
0.019$\times$0.017 (56$^{\circ}$) \\ 
\enddata

\tablecomments{
Col.(1): Frequency range (in GHz) used to extract continuum emission.
Frequencies of obvious emission and absorption lines are excluded.
Col.(2): Flux (in mJy beam$^{-1}$) at the emission peak.
The value at the highest flux pixel (1.5 mas pixel$^{-1}$) is adopted.
The detection significance relative to the root mean square (rms) noise
is shown in parentheses.
Possible systematic uncertainties coming from absolute flux
calibration ambiguity in individual ALMA observations and
the choice of frequency range to determine the continuum level
are not taken into account.
Col.(3): Coordinate of the continuum emission peak in ICRS.
Col.(4): The rms noise level (1$\sigma$) (in mJy beam$^{-1}$), derived
from the standard deviation of sky signals.
Col.(5): Synthesized beam (in arcsec $\times$ arcsec) and position angle
(in degrees).
The position angle is 0$^{\circ}$ along the north-south direction
and increases in the counterclockwise direction.
}

\end{deluxetable*}

We define the locations of the HCN J=4--3 emission peaks in the
western and eastern torus in Figure \ref{fig:mom0}a as ``W-peak'' and
``E-peak'', respectively, and use these peaks for the following
discussion for two reasons: (1) the HCN emission is brighter than the
HCO$^{+}$ emission, and (2) the newly taken J=4--3 data have a smaller
synthesized beam size than the reanalyzed J=3--2 data, both of which
make the HCN J=4--3 data the best for determining the peak position in
the most accurate manner. 
The coordinates of the peak positions of the HCN J=4--3 and other
emission lines are summarized in Table \ref{tab:peak}.

\begin{deluxetable*}{cccccc}
\tabletypesize{\scriptsize}
\tablecaption{Molecular Emission Line Peak Position \label{tab:peak}} 
\tablewidth{0pt}
\tablehead{
\colhead{Line} & \colhead{Torus position} & \colhead{Peak flux} & 
\colhead{Peak coordinate} & \colhead{rms} & \colhead{Synthesized beam} \\
\colhead{} & \colhead{} & \colhead{(Jy km s$^{-1}$ beam$^{-1}$)} &  
\colhead{(RA,DEC)ICRS} & \colhead{(Jy km s$^{-1}$ beam$^{-1}$)} & 
\colhead{($''$ $\times$ $''$) ($^{\circ}$)} \\  
\colhead{(1)} & \colhead{(2)} & \colhead{(3)}  & \colhead{(4)}  &
\colhead{(5)} & \colhead{(6)}
}
\startdata 
HCN J=4--3 & W-torus & 0.77 (13.9$\sigma$) & 
(02$^{\rm h}$ 42$^{\rm m}$ 40.7084$^{\rm s}$, $-$00$^{\circ}$ 00$'$
47$\farcs$940) (W-peak) & 0.056 & 0.014$\times$0.013 ($-$83$^{\circ}$) \\
 & E-torus & 0.63 (11.4$\sigma$) & 
(02$^{\rm h}$ 42$^{\rm m}$ 40.710$^{\rm s}$, $-$00$^{\circ}$ 00$'$
47$\farcs$946) (E-peak) &  & \\
 & C-abs & $-$0.14 ($-$2.6$\sigma$) & 
(02$^{\rm h}$ 42$^{\rm m}$ 40.709$^{\rm s}$, $-$00$^{\circ}$ 00$'$
47$\farcs$949) &  & \\
HCO$^{+}$ J=4--3 & W-torus & 0.68 (11.8$\sigma$) & 
(02$^{\rm h}$ 42$^{\rm m}$ 40.709$^{\rm s}$, $-$00$^{\circ}$ 00$'$
47$\farcs$939) & 0.058 & 0.014$\times$0.013 ($-$83$^{\circ}$) \\
 & E-torus & 0.41 (7.1$\sigma$) & 
(02$^{\rm h}$ 42$^{\rm m}$ 40.710$^{\rm s}$, $-$00$^{\circ}$ 00$'$
47$\farcs$946) & & \\
 & C-abs & $-$0.25 ($-$4.3$\sigma$) & 
(02$^{\rm h}$ 42$^{\rm m}$ 40.709$^{\rm s}$, $-$00$^{\circ}$ 00$'$
47$\farcs$951) & & \\
HCN J=3--2 & W-torus & 0.44 (11.7$\sigma$) & 
(02$^{\rm h}$ 42$^{\rm m}$ 40.708$^{\rm s}$, $-$00$^{\circ}$ 00$'$
47$\farcs$940) & 0.037 & 0.020$\times$0.018 (54$^{\circ}$) \\
 & E-torus & 0.25 (6.8$\sigma$) & 
(02$^{\rm h}$ 42$^{\rm m}$ 40.710$^{\rm s}$, $-$00$^{\circ}$ 00$'$
47$\farcs$950) & & \\
 & C-abs & $-$0.18 ($-$4.9$\sigma$) & 
(02$^{\rm h}$ 42$^{\rm m}$ 40.709$^{\rm s}$, $-$00$^{\circ}$ 00$'$
47$\farcs$949) & & \\
HCO$^{+}$ J=3--2 & W-torus & 0.34 (10.8$\sigma$) & 
(02$^{\rm h}$ 42$^{\rm m}$ 40.708$^{\rm s}$, $-$00$^{\circ}$ 00$'$
47$\farcs$940) & 0.032 & 0.020$\times$0.018 (55$^{\circ}$) \\
 & E-torus & 0.30 (9.4$\sigma$) & 
(02$^{\rm h}$ 42$^{\rm m}$ 40.710$^{\rm s}$, $-$00$^{\circ}$ 00$'$
47$\farcs$945) & & \\
 & C-abs & $-$0.22 ($-$6.9$\sigma$) & 
(02$^{\rm h}$ 42$^{\rm m}$ 40.709$^{\rm s}$, $-$00$^{\circ}$ 00$'$
47$\farcs$947) & & \\
CS J=7--6 & W-torus & 0.072 (5.8$\sigma$) & 
(02$^{\rm h}$ 42$^{\rm m}$ 40.709$^{\rm s}$, $-$00$^{\circ}$ 00$'$
47$\farcs$937) & 0.012 & 0.014$\times$0.013 ($-$82$^{\circ}$) 
\enddata

\tablecomments{
Col.(1): Molecular line.
Col.(2): Western torus (W-torus), eastern torus (E-torus), or
absorption peak near the continuum emission peak position (C-abs).
Col.(3): Flux (in Jy km s$^{-1}$ beam$^{-1}$) at the emission or
absorption peak.
The value at the highest or most negative flux pixel (1.5 mas
pixel$^{-1}$) in the integrated intensity (moment 0) map in Figure
\ref{fig:mom0} or \ref{fig:CS76}a is adopted. 
The detection significance relative to the rms noise is shown in
parentheses.
Possible systematic uncertainties coming from absolute flux
calibration ambiguity in individual ALMA observations and
the choice of frequency range used to determine the continuum level
are not taken into account.
Col.(4): Coordinate of the emission or absorption peak in ICRS.
Col.(5): The rms noise level (1$\sigma$) (in Jy km s$^{-1}$
beam$^{-1}$), derived from the standard deviation of sky signals.
Col.(6): Synthesized beam (in arcsec $\times$ arcsec) and position angle
(in degrees).
The definition of the position angle is the same as that in Table
\ref{tab:cont}.
}

\end{deluxetable*}

\begin{figure*}[!hbt]
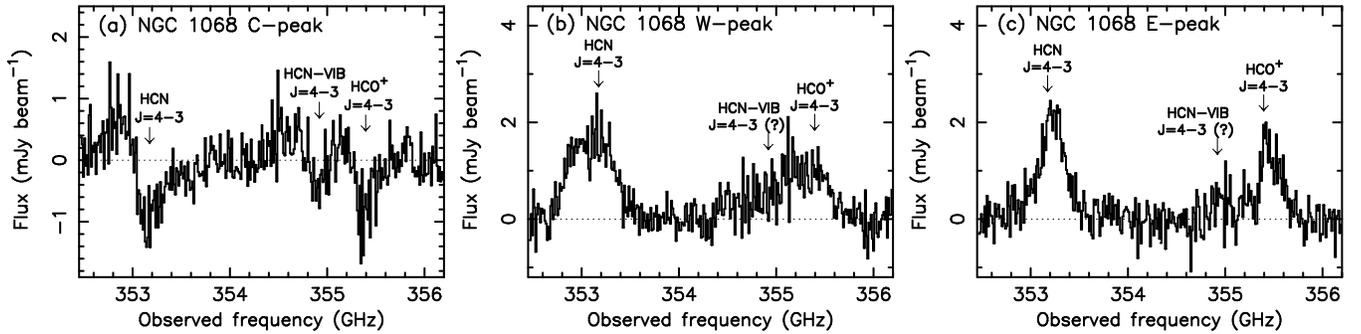

\begin{center}
\hspace{-0.4cm}
\includegraphics[scale=0.26,angle=-90]{f2a.eps} 
\includegraphics[scale=0.26,angle=-90]{f2b.eps} 
\includegraphics[scale=0.26,angle=-90]{f2c.eps} 
\end{center}
\vspace{+0.7cm}
\caption{
Observed beam-sized spectra at (a) $\sim$350 GHz continuum emission peak
(02$^{\rm h}$ 42$^{\rm m}$ 40.709$^{\rm s}$, $-$00$^{\circ}$ 00$'$
47$^{\prime\prime}.946$)ICRS (defined as "C-peak"),
(b) HCN J=4--3 emission peak in the western torus
(02$^{\rm h}$ 42$^{\rm m}$ 40.7084$^{\rm s}$, $-$00$^{\circ}$ 00$'$
47$^{\prime\prime}.940$)ICRS (defined as "W-peak"),
and (c) HCN J=4--3 emission peak in the eastern torus
(02$^{\rm h}$ 42$^{\rm m}$ 40.710$^{\rm s}$, $-$00$^{\circ}$ 00$'$
47$^{\prime\prime}.946$)ICRS (defined as "E-peak").
The abscissa is the observed frequency (in GHz), and 
the ordinate is the flux density (in mJy beam$^{-1}$).
The beam size is 0$\farcs$014 $\times$ 0$\farcs$013 
(0.98 pc $\times$ 0.91 pc).
Downward arrows indicate the expected frequency of HCN J=4--3
($\nu_{\rm rest}$ = 354.505 GHz),
HCO$^{+}$ J=4--3 ($\nu_{\rm rest}$ = 356.734 GHz), and HCN-VIB J=4--3
($\nu_{\rm rest}$ = 356.256 GHz) lines for the systemic optical LSR
velocity of NGC 1068, V$_{\rm sys}$ = 1130 km s$^{-1}$
\citep[e.g.,][]{ima16a,ima18a,imp19,ima20,gal24}.
The horizontal dotted straight line indicates the zero-flux level.
\label{fig:spectraPeak}
}
\end{figure*}

Figure \ref{fig:spectraPeak} displays beam-sized spectra at the
continuum emission peak (C-peak) and at the HCN J=4--3 emission peaks in the
western and eastern torus (W-peak and E-peak, respectively).
At the C-peak, the HCN J=4--3 and HCO$^{+}$ J=4--3 lines are observed in
absorption (Figure \ref{fig:spectraPeak}a).
At the W-peak and E-peak, the HCN J=4--3 and HCO$^{+}$ J=4--3 emission
lines are clearly detected (Figure \ref{fig:spectraPeak}b,c).

\begin{figure*}[!hbt]
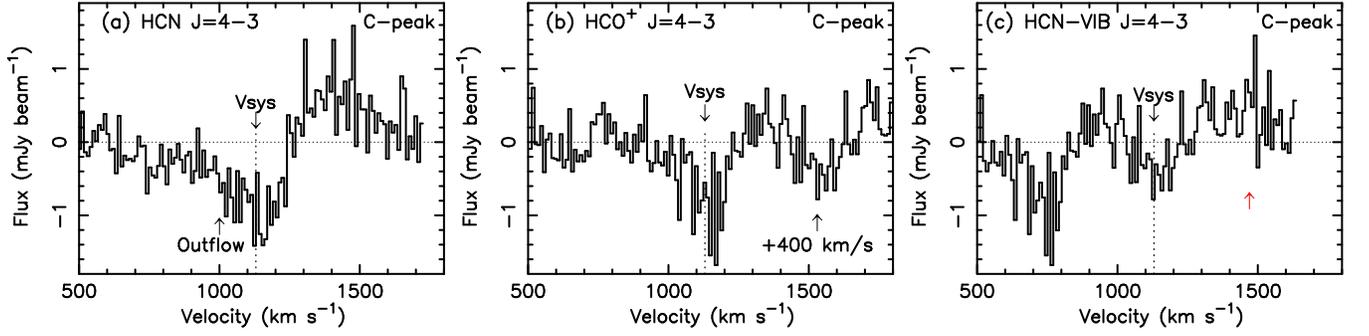

\begin{center}
\hspace{-0.4cm}
\includegraphics[scale=0.26,angle=-90]{f3a.eps} 
\includegraphics[scale=0.26,angle=-90]{f3b.eps} 
\includegraphics[scale=0.26,angle=-90]{f3c.eps} 
\end{center}
\vspace{+0.7cm}
\caption{
Velocity profile of (a) HCN J=4--3, (b) HCO$^{+}$ J=4--3,
and (c) HCN-VIB J=4--3 line in the beam-sized spectrum toward the
$\sim$350 GHz continuum emission peak ("C-peak").
The abscissa is the optical LSR velocity (in km s$^{-1}$), and the ordinate
is the flux density (in mJy beam$^{-1}$).
A downward arrow is added at the systemic velocity of NGC 1068
(V$_{\rm sys}$ = 1130 km s$^{-1}$), labeled as
"V$_{\rm sys}$".
The horizontal thin dotted straight line indicates the zero-flux level.
In (b), an upward arrow is added at the velocity of
V = 1530 km s$^{-1}$ (V$_{\rm sys}$ $+$ 400 km s$^{-1}$) to indicate
an absorption feature that is 400 km s$^{-1}$ redshifted relative to the systemic
velocity.
In (c), a red upward arrow is added at the velocity of V = 1470
km s$^{-1}$ (V$_{\rm sys}$ $+$ 340 km s$^{-1}$)
to indicate redshifted emission at the innermost ($\lesssim$1 pc)
western torus (see $\S$3).
\label{fig:spectraPeakVopt}
}
\end{figure*}

The velocity profiles of the HCN J=4--3 and HCO$^{+}$ J=4--3 lines
toward the C-peak in the beam-sized spectrum are shown in Figure
\ref{fig:spectraPeakVopt}.
Absorption features at the systemic velocity
(V$_{\rm sys}$ = 1130 km s$^{-1}$) are clearly detected for both lines
\footnote{
\citet{gam25} mentioned that this absorption feature at the systemic
velocity was not detected for HCO$^{+}$ J=4--3 in their analysis using
their data with comparable beam size ($\sim$0$\farcs$016 or $\sim$1.1
pc) to ours. We detect this absorption feature for HCO$^{+}$ J=4--3 in
our data using the analysis method described in $\S$2, as in the case
for HCO$^{+}$ J=3--2 \citep{ima20}.   
}
at the C-peak (Figures \ref{fig:spectraPeak}a and \ref{fig:spectraPeakVopt}a,b),
as previously seen for the J=3--2 lines of HCN and HCO$^{+}$ at the
$\sim$260 GHz continuum emission peak in $\sim$1.5 pc ($\sim$0$\farcs$02)
beam-sized spectra \citep[e.g.,][]{imp19,ima20}.

For HCO$^{+}$ J=4--3, an absorption feature is seen at $\sim$400 km
s$^{-1}$ redshifted with respect to the systemic velocity (Figure
\ref{fig:spectraPeakVopt}b).
The $\sim$400 km s$^{-1}$ redshifted HCO$^{+}$ J=4--3 absorption
feature was also reported by \citet{gam25}, who interpreted it as due to a
highly redshifted ($\sim$400 km s$^{-1}$) infalling gas clump toward
the continuum-emitting mass-accreting SMBH, without strong interaction
with the torus.
However, we note that the observed frequency spectrally coincides with
the frequency at the systemic velocity (V$_{\rm sys}$ = 1130 km s$^{-1}$),
of the HCN-VIB J=4--3 line ($\nu{\rm rest}$ = 356.256 GHz),
which is $\sim$400 km s$^{-1}$ redshifted relative to the HCO$^{+}$
J=4--3 line.
We discuss these two scenarios in Appendix A.

For HCN J=4--3, a broad absorption tail on the blueshifted
side (V = 700--1100 km s$^{-1}$ or V$_{\rm sys}$ $-$ [30--430] km
s$^{-1}$) is seen (Figure \ref{fig:spectraPeakVopt}a).
The same absorption tail was previously detected for HCN J=3--2 in the
$\sim$1.5 pc ($\sim$0$\farcs$02) beam-sized spectrum toward the
$\sim$260 GHz continuum emission peak and was interpreted as
originating from a nuclear outflow \citep{imp19,ima20}.
The corresponding absorption tail is less clear in HCO$^{+}$ J=4--3
(Figure \ref{fig:spectraPeakVopt}b) and HCO$^{+}$ J=3--2 \citep{ima20}.

\begin{figure*}[!hbt]
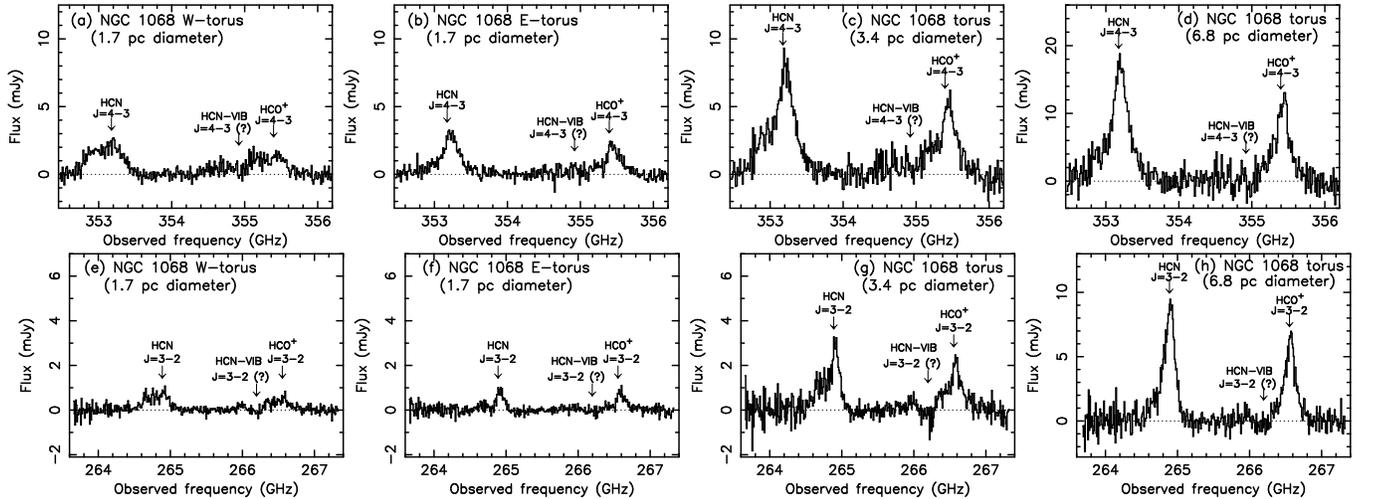

\begin{center}
\hspace{-0.4cm}
\includegraphics[scale=0.195,angle=-90]{f4a.eps} 
\includegraphics[scale=0.195,angle=-90]{f4b.eps} 
\includegraphics[scale=0.195,angle=-90]{f4c.eps}
\includegraphics[scale=0.195,angle=-90]{f4d.eps} \\
\includegraphics[scale=0.195,angle=-90]{f4e.eps} 
\includegraphics[scale=0.195,angle=-90]{f4f.eps} 
\includegraphics[scale=0.195,angle=-90]{f4g.eps} 
\includegraphics[scale=0.195,angle=-90]{f4h.eps} 
\end{center}
\vspace{+0.5cm}
\caption{
{\it (Upper)}: Observed area-integrated spectra of HCN J=4--3 and
HCO$^{+}$ J=4--3, extracted with
(a) a 1.7 pc (24 mas) diameter circular aperture 
centered at the W-peak,
(b) a 1.7 pc diameter circular aperture centered at the E-peak,
(c) a 3.4 pc (48 mas) diameter circular aperture centered at the C-peak,
and 
(d) a 6.8 pc (96 mas) diameter circular aperture centered at the C-peak.
{\it (Lower)}: Observed area-integrated spectra of HCN J=3--2 and
HCO$^{+}$ J=3--2, extracted with
(e) a 1.7 pc (24 mas) diameter
circular 
aperture centered at the W-peak,
(f) a 1.7 pc diameter circular aperture centered at the E-peak,
(g) a 3.4 pc (48 mas) diameter circular aperture centered at the C-peak, and
(h) a 6.8 pc (96 mas) diameter circular aperture centered at the C-peak.
The abscissa is the observed frequency (in GHz), and the ordinate is the flux
density (in mJy).
Downward arrows are added at the expected frequencies of the HCN, HCO$^{+}$,
and HCN-VIB lines at J=4--3 and J=3--2, corresponding to the systemic 
velocity of V$_{\rm sys}$ = 1130 km s$^{-1}$.
The horizontal dotted straight line indicates the zero-flux level.
\label{fig:spectraArea}
}
\end{figure*}

Area-integrated spectra of the western and eastern torus, obtained
with a 1.7 pc (24 mas)  
diameter circle around the W-peak and E-peak, and spectra that include
both sides of the tori,  
obtained with 3.4 pc (48 mas) and 6.8 pc (96 mas) diameter circles
around the C-peak, are shown in Figure \ref{fig:spectraArea}.
For the HCN emission, an excess emission tail is more significantly
recognizable on the lower-frequency side of the main emission
component at the systemic velocity in the western torus
(Figure \ref{fig:spectraArea}a,e) than in the eastern torus
(Figure \ref{fig:spectraArea}b,f).
Gaussian fits are applied to the detected HCN and HCO$^{+}$ emission
lines in the area-integrated spectra (Appendix B).
Table \ref{tab:Gauss} summarizes the estimated emission line fluxes.

\begin{deluxetable*}{ccccc}[!hbt]
\tabletypesize{\scriptsize}
\tablecaption{Molecular Emission Line Flux \label{tab:Gauss}} 
\tablewidth{0pt}
\tablehead{
\colhead{Torus position} & \colhead{HCN J=4--3} & \colhead{HCO$^{+}$ J=4--3} & 
\colhead{HCN J=3--2} & \colhead{HCO$^{+}$ J=3--2} \\
\colhead{} & \colhead{(Jy km s$^{-1}$)} & \colhead{(Jy km s$^{-1}$)} & 
\colhead{(Jy km s$^{-1}$)} &  \colhead{(Jy km s$^{-1}$)} \\
\colhead{(1)} & \colhead{(2)} & \colhead{(3)}  & \colhead{(4)}  &
\colhead{(5)} 
}
\startdata 
W-torus (1.7 pc diameter) & 0.96$\pm$0.10 & 0.60$\pm$0.10 &
0.31$\pm$0.07 & 0.18$\pm$0.02 \\  
E-torus (1.7 pc diameter) & 0.73$\pm$0.05 & 0.47$\pm$0.06 &
0.17$\pm$0.02 & 0.14$\pm$0.02 \\
Torus (3.4 pc diameter) & 2.5$\pm$0.2 & 1.5$\pm$0.1 & 0.76$\pm$0.15 &
0.54$\pm$0.05 \\
Torus (6.8 pc diameter) & 4.3$\pm$0.3 & 2.5$\pm$0.2 & 2.0$\pm$0.1 &
1.4$\pm$0.1 \\
\enddata

\tablecomments{
Col.(1): Torus region.
Cols.(2)--(5): Gaussian-fit velocity-integrated emission line fluxes of
(2) HCN J=4--3, (3) HCO$^{+}$ J=4--3, (4) HCN J=3--2, and (5)
HCO$^{+}$ J=3--2, in units of Jy km s$^{-1}$.}

\end{deluxetable*}

The CS J=7--6 ($\nu_{\rm rest}$ = 342.883 GHz) emission line is also
marginally detected in the NGC 1068 compact ($\lesssim$3--5 pc) torus
region.
The details are presented in Appendix C.

Figure \ref{fig:mom1} displays intensity-weighted mean velocity
(moment 1) maps of the J=4--3 lines of HCN and HCO$^{+}$, together with
those of the reanalyzed J=3--2 lines of HCN and HCO$^{+}$.
Figures \ref{fig:mom1}a--d highlight the
redshifted (V $\gtrsim$ 1330 km s$^{-1}$ or $\Delta$V$_{\rm red}$ $\gtrsim$
$+$200 km s$^{-1}$) emission component in the innermost
($\lesssim$1 pc) western torus relative to the systemic velocity
(V$_{\rm sys}$ = 1130 km s$^{-1}$), which was previously seen for HCN J=3--2
\citep[e.g.,][]{imp19,ima20} (see also Figure \ref{fig:mom1}c).
This redshifted (V $\gtrsim$ 1330 km s$^{-1}$) component in the
innermost ($\lesssim$1 pc) western torus is clearly detected in the
new HCN J=4--3 data as well (Figure \ref{fig:mom1}a).

\begin{figure*}[!hbt]
\begin{center}
\hspace{-0.4cm}
\includegraphics[scale=0.37]{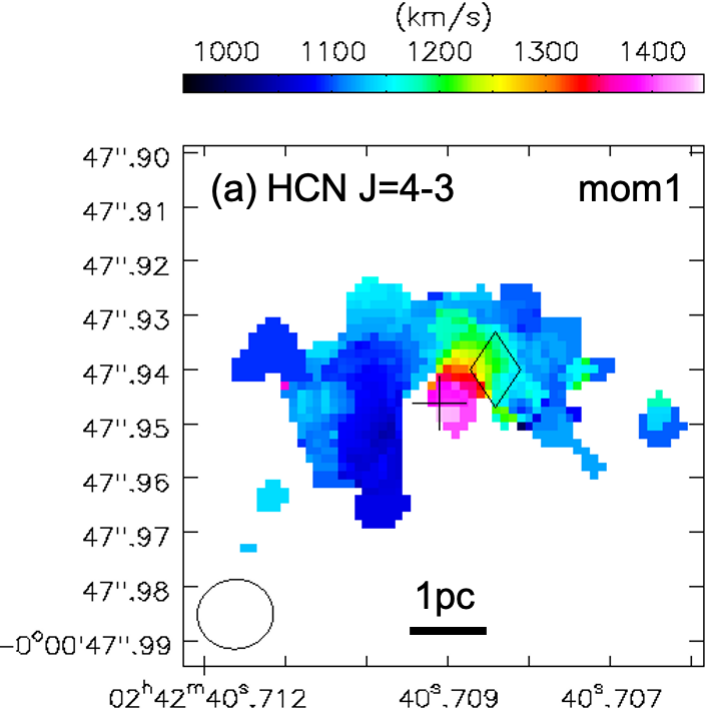} 
\includegraphics[scale=0.37]{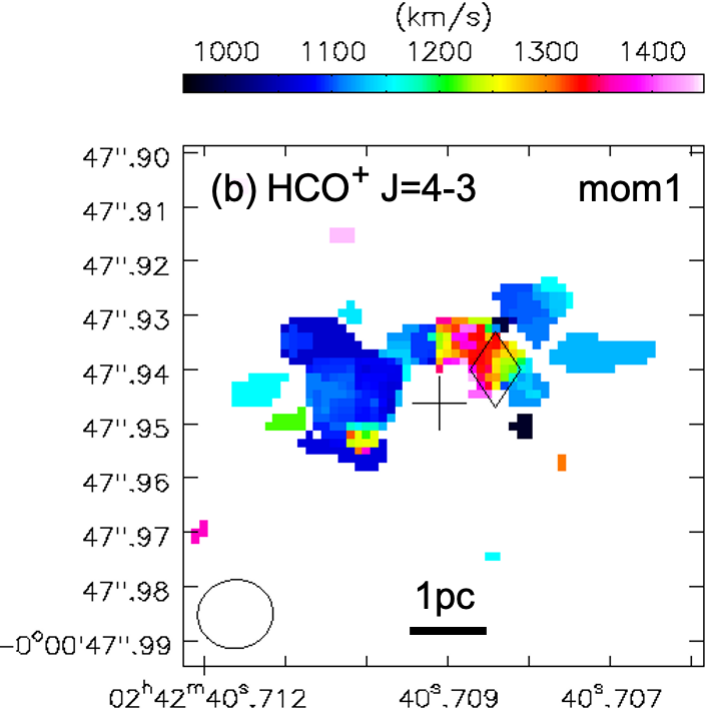} 
\includegraphics[scale=0.37]{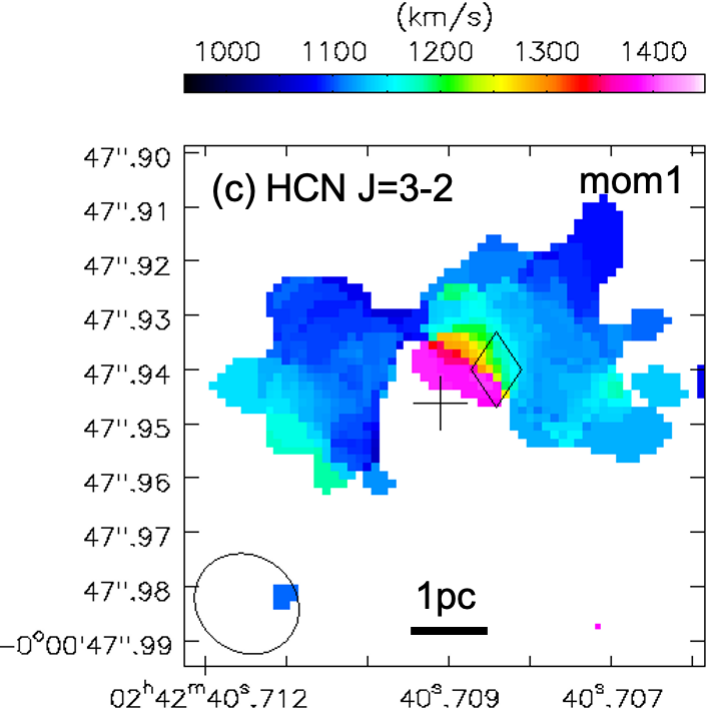} 
\includegraphics[scale=0.37]{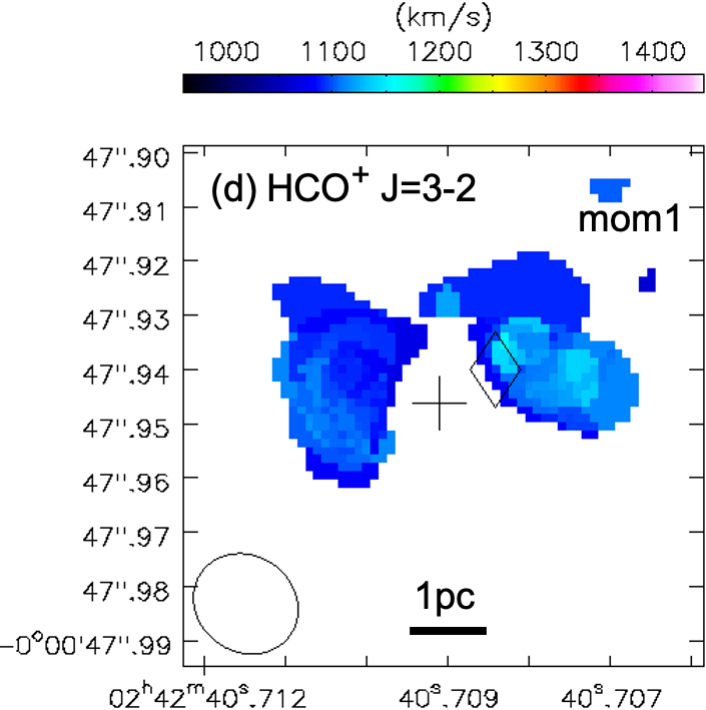} \\
\hspace*{-9.5cm}
\includegraphics[scale=0.37]{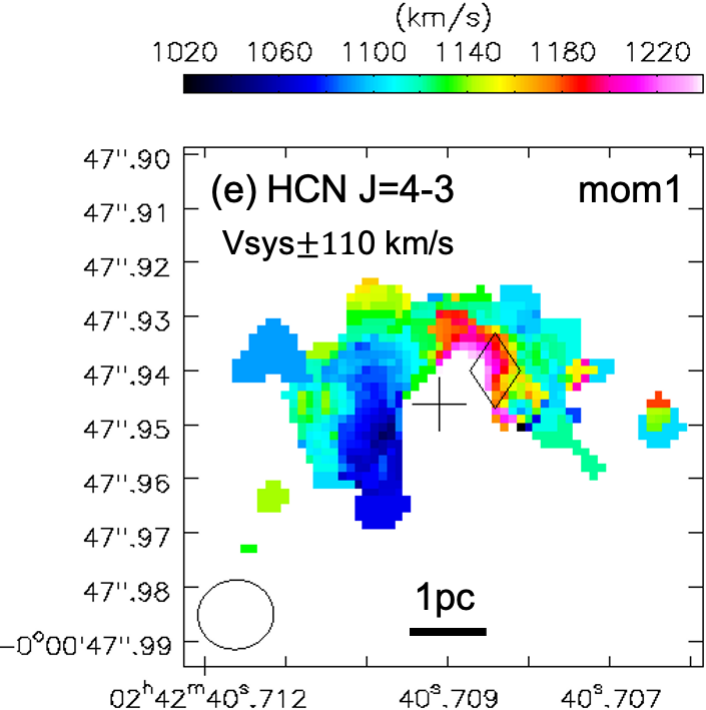} 
\includegraphics[scale=0.37]{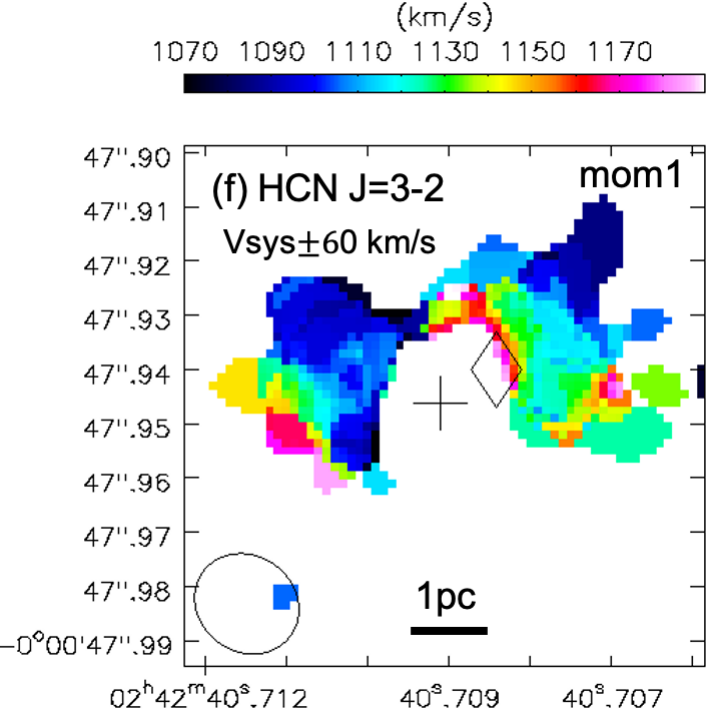} 
\end{center}
\vspace{-0.2cm}
\caption{
Intensity-weighted mean velocity (moment 1) map of
(a) HCN J=4--3, (b) HCO$^{+}$ J=4--3, (c) HCN J=3--2, and
(d) HCO$^{+}$ J=4--3, created with 1.5 mas pixel$^{-1}$.
The displayed velocity range is 1130$^{+320}_{-170}$ km s$^{-1}$
(960--1450 km s$^{-1}$) for (a)-(d), to clearly display the innermost
($\lesssim$1 pc) high-velocity redshifted HCN emission in the western
torus \citep[e.g.,][]{imp19,ima20} (see text in $\S$3).
Moment 1 maps of (e) HCN J=4--3 and (f) HCN J=3--2 are shown with the
velocity ranges of 1130$\pm$110 km s$^{-1}$ (= 1020--1240 km s$^{-1}$)
and 1130$\pm$60 km s$^{-1}$ (= 1070--1190 km s$^{-1}$), respectively,
to display overall redshifted and blueshifted motion of the torus
dense molecular gas emission with respect to the systemic velocity
(V$_{\rm sys}$ = 1130 km s$^{-1}$).
The plus mark is the $\sim$350 GHz continuum emission peak position
(02$^{\rm h}$ 42$^{\rm m}$ 40.709$^{\rm s}$, $-$00$^{\circ}$ 00$'$
47$\farcs$946)ICRS defined as C-peak.
The open diamond mark is the emission peak in the western torus in the
HCN J=4--3 moment 0 map 
(02$^{\rm h}$ 42$^{\rm m}$ 40.7084$^{\rm s}$, $-$00$^{\circ}$ 00$'$
47$\farcs$940)ICRS defined as W-peak.
The horizontal thick black bar at the bottom center of each panel
corresponds to 1 pc at the distance of NGC 1068.
The open circle at the lower-left part of each panel indicates
the synthesized beam size, which is 0$\farcs$014 $\times$ 0$\farcs$013
($\lesssim$1 pc) for the J=4--3 lines of HCN and HCO$^{+}$, and
0$\farcs$020 $\times$ 0$\farcs$018 for the J=3--2 lines.
An appropriate cutoff ($\sim$4.5$\sigma$) is applied to all the moment
1 maps to prevent them from being dominated by noise.
\label{fig:mom1}
}
\end{figure*}

The redshifted ($\Delta$V$_{\rm red}$ $\gtrsim$ $+$200 km s$^{-1}$) 
HCN emission component at the innermost ($\lesssim$1 pc) western 
torus can contribute to the 
excess HCN J=4--3 and HCN J=3--2 emission tails on the 
lower-frequency (redshifted) 
side of the main emission components at the systemic velocity, 
which are clearly seen 
in the area-integrated spectra at the western torus 
(Figure \ref{fig:spectraArea}a,e).
This redshifted emission component is not clearly seen 
in the moment 1 maps of 
the HCO$^{+}$ J=4--3 and HCO$^{+}$ J=3--2 lines at the 
innermost western torus 
(Figure \ref{fig:mom1}b,d).
We will use the HCN J=4--3 line emission detected in 
our $\lesssim$1 pc resolution 
data to investigate the innermost dense molecular gas kinematics 
in the western torus.

\begin{figure}[!hbt]
\vspace*{-1.0cm}
\begin{center}
\includegraphics[scale=0.37,angle=-90]{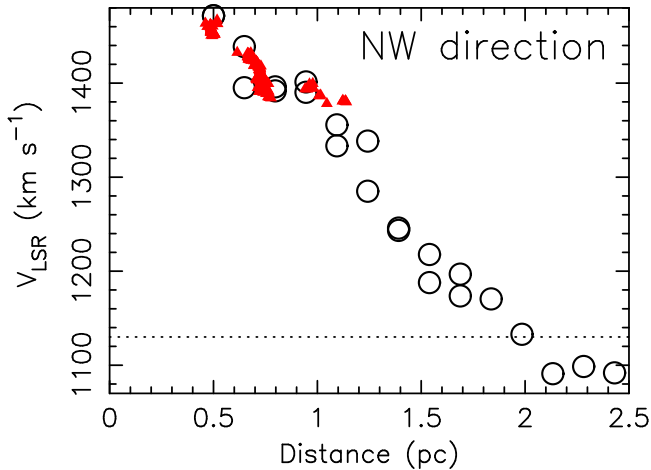} 
\end{center}
\vspace{+1.0cm}
\caption{
Intensity-weighted velocity of the HCN J=4--3 (open circles)
and 22 GHz (1.4 cm) H$_{2}$O maser emission (red filled triangles)
(ordinate) as a function of distance from the SMBH position (continuum
emission peak) along the north-western direction (PA = 135$^{\circ}$)
(abscissa). 
We shift our ALMA HCN J=4--3 data by $\sim$9 mas ($\sim$0.65 pc)
toward the north-western direction (see also \citet{gam25}),
assuming that the most highly redshifted (V $\sim$ 1470 km s$^{-1}$ or
$\Delta$V$_{\rm red}$ $\sim$ $+$340 km s$^{-1}$) HCN J=4--3 emission
component is at $\sim$0.5 pc from the VLBI 22 GHz continuum emission
peak, where the 22 GHz H$_{2}$O maser emission also displays the most
redshifted components with V $\sim$ 1470 km s$^{-1}$
($\Delta$V$_{\rm red}$ $\sim$ $+$340 km s$^{-1}$) \citep{gal23}.
This is because the continuum emission peak positions often do not
perfectly agree between ALMA at $\sim$260--350 GHz and VLBI at
$\sim$22 GHz, possibly due to non-perfect astrometric accuracy.
Filled circles: 22 GHz H$_{2}$O maser emission.
Data in the R1, R2, and R3 regions, and those with $>$1450 km s$^{-1}$
in the R4 region \citep{gal23} are plotted.
Distance in the abscissa is calculated from the offset values
from the S1 in Table 1 of \citet{gal23}.
Open stars: HCN J=4--3 emission with 2-pixel width.
The horizontal dotted straight line indicates the systemic velocity of
V$_{\rm sys}$ = 1130 km s$^{-1}$.
The counter-rotation signature, with a velocity change from redshifted
at $\lesssim$2 pc to blueshifted at $\gtrsim$2 pc, is seen along the
north-western direction.
\label{fig:Wtorus}
}
\end{figure}

VLBI observations at centimeter wavelengths with very high angular 
resolution ($\lesssim$2 mas) 
show that NGC 1068 has a rotating 22 GHz (1.4 cm) H$_{2}$O maser 
emitting disk \citep[e.g.,][]{gre96,gre97,gal04,mor23,gal23,gal24}.
A larger number of bright, redshifted 22 GHz (1.4 cm) H$_{2}$O 
maser spots are detected at $\sim$0.5--1.2 pc ($\sim$7--17 mas) 
on the north-western (NW) side of the central mass-accreting SMBH 
(the VLBI radio 22 GHz continuum emission peak, named ``S1'')  
than on the blueshifted south-eastern (SE) side 
\citep[e.g.,][]{gre97,gal04,mor23,gal23,gal24}.
The most redshifted H$_{2}$O maser emission has an intensity-weighted 
velocity of 
V $\sim$ 1470 km s$^{-1}$ ($\Delta$V$_{\rm red}$ $\sim$ $+$340 km s$^{-1}$)  
at $\sim$0.5 pc north-west of the SMBH \citep{gal23}.
Allowing for a slight ($\sim$9 mas) positional shift between the 
ALMA $\sim$350 GHz and VLBI $\sim$22 GHz data, possibly caused by
astrometric uncertainties in both observations \citep{gam25} 
\footnote{
According to the ALMA Cycle 9 Proposer's Guide, the best
absolute astrometric accuracy is $\sim$10\% of the synthesized beam,
namely $\lesssim$2 mas for our HCN J=4--3 data.
\citet{gam25} adopted an even larger positional shift of $\sim$25 mas
between their ALMA and VLBI data than ours ($\sim$9 mas). 
},
Figure \ref{fig:Wtorus} compares the
velocity profiles of the VLBI-detected 22 GHz H$_{2}$O maser emitting
spots and the ALMA-detected HCN J=4--3 emission.
In the HCN J=4--3 data, the most redshifted component on the NW side has 
an intensity-weighted velocity as high as V $\sim$ 1470 km s$^{-1}$ 
($\Delta$V$_{\rm red}$ $\sim$ $+$340 km s$^{-1}$) (Figure \ref{fig:Wtorus}).
The rough agreement of the velocity profiles of the redshifted 22 GHz
H$_{2}$O maser and the HCN J=4--3 emission at $\sim$0.5--1.2 pc along
the NW direction (after the slight positional shift) in Figure
\ref{fig:Wtorus} suggests that the redshifted HCN J=4--3 line-emitting
dense molecular gas in the innermost ($\lesssim$1 pc) western torus is
physically connected to the VLBI-detected redshifted H$_{2}$O maser  
emitting rotating disk.

Figures \ref{fig:mom1}e and \ref{fig:mom1}f present
intensity-weighted mean velocity (moment 1) maps of the HCN J=4--3 and
HCN J=3--2 emission using the same redshifted and blueshifted velocity
ranges relative to the systemic velocity(V$_{\rm sys}$ ± 110 km s$^{-1}$ 
and V$_{\rm sys}$ ± 60 km s$^{-1}$,
respectively) to better visualize the overall rotation patterns of the
compact ($\lesssim$3--5 pc) HCN emission.
In Figure \ref{fig:mom1}f, the HCN J=3--2 emission in the western
torus is redshifted in the inner part ($\lesssim$2 pc) and blueshifted
in the outer ($\gtrsim$2 pc) part, while in the
eastern torus it is blueshifted in the inner part ($\lesssim$2 pc)
and redshifted in the outer ($\gtrsim$2 pc) part, i.e.,
they appear to be counter-rotating \citep{imp19,ima20}.
For the HCN J=4--3 line, a similar apparent counter-rotation is seen
in the western torus (Figure \ref{fig:mom1}e), but in the eastern
torus only the inner ($\lesssim$2 pc) blueshifted emission component
is clearly visible, without a significant outer ($\gtrsim$2 pc)
redshifted emission component.
This suggests that molecular gas at the outer ($\gtrsim$2 pc) eastern
torus is not sufficiently warm and dense, because the upper energy
level (E$_{\rm upper}$) and critical density (n$_{\rm crit}$) for HCN
J=4--3 (E$_{\rm upper}$ $\sim$ 43 K and n$_{\rm crit}$ $\sim$ 1
$\times$ 10$^{7}$ cm$^{-3}$ at 50--100 K) are higher than those for HCN
J=3--2 (E$_{\rm upper}$ $\sim$ 26 K and n$_{\rm crit}$ $\sim$ 4--6
$\times$ 10$^{6}$ cm$^{-3}$).

In Figure \ref{fig:Wtorus}, the velocity changes from an inner
redshift ($\lesssim$2 pc) to an outer blueshift ($\gtrsim$2 pc) 
along the north-western direction for HCN J=4--3, but the outer
blueshifted velocity is only 
$\Delta$V$_{\rm blue}$ $\sim$ $-$40 km s$^{-1}$ relative to the
systemic velocity of NGC 1068 (V$_{\rm sys}$ = 1130 km s$^{-1}$).
This is much slower than the Keplerian rotation velocity of
$\sim$130--150 km s$^{-1}$ at $\sim$2--2.5 pc for the central SMBH
mass of NGC 1068 with M$_{\rm SMBH}$ $\sim$ 1 $\times$ 10$^{7}$M$_{\odot}$
\citep[e.g.,][]{gre96,hur02,lod03,gal23,gal24}, as has already
been noted in previous studies \citep[e.g.,][]{ima20,vol22,gam25}.

\begin{figure*}[!hbt]
\begin{center}
\includegraphics[scale=0.44]{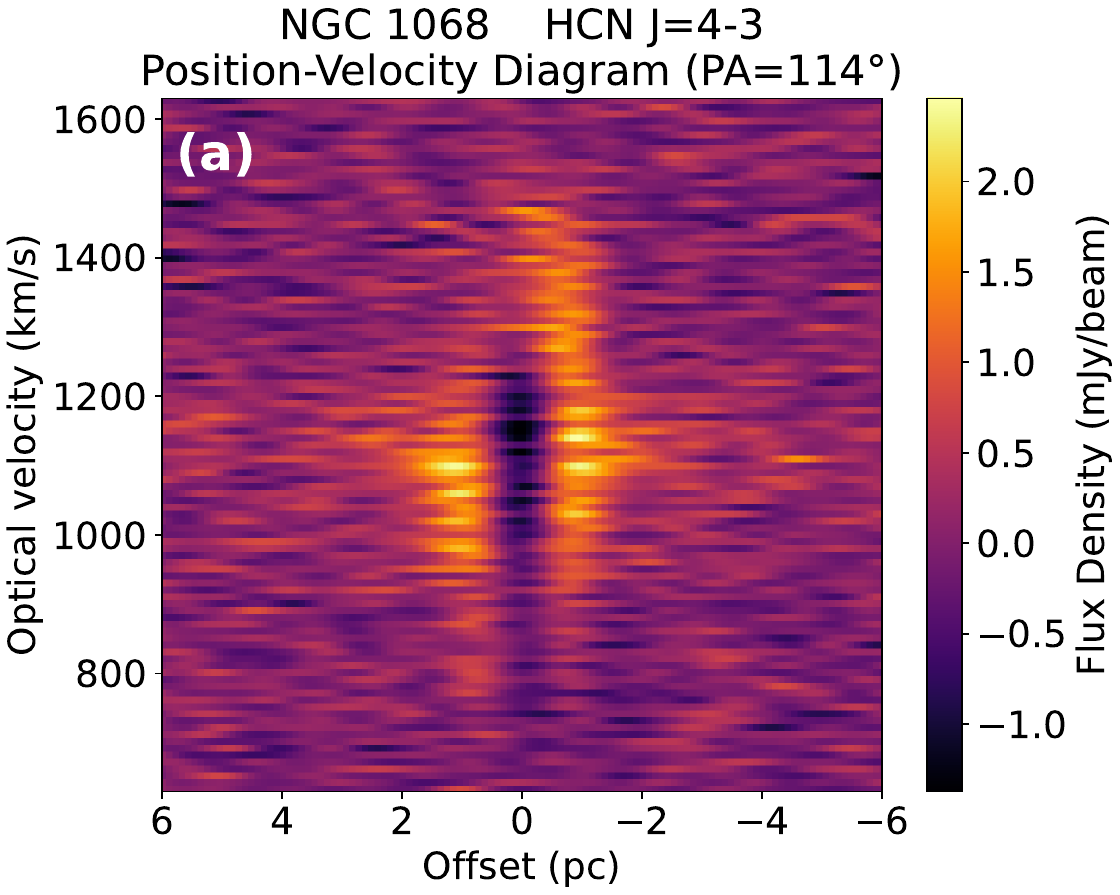} 
\hspace{1.0cm}
\includegraphics[scale=0.44]{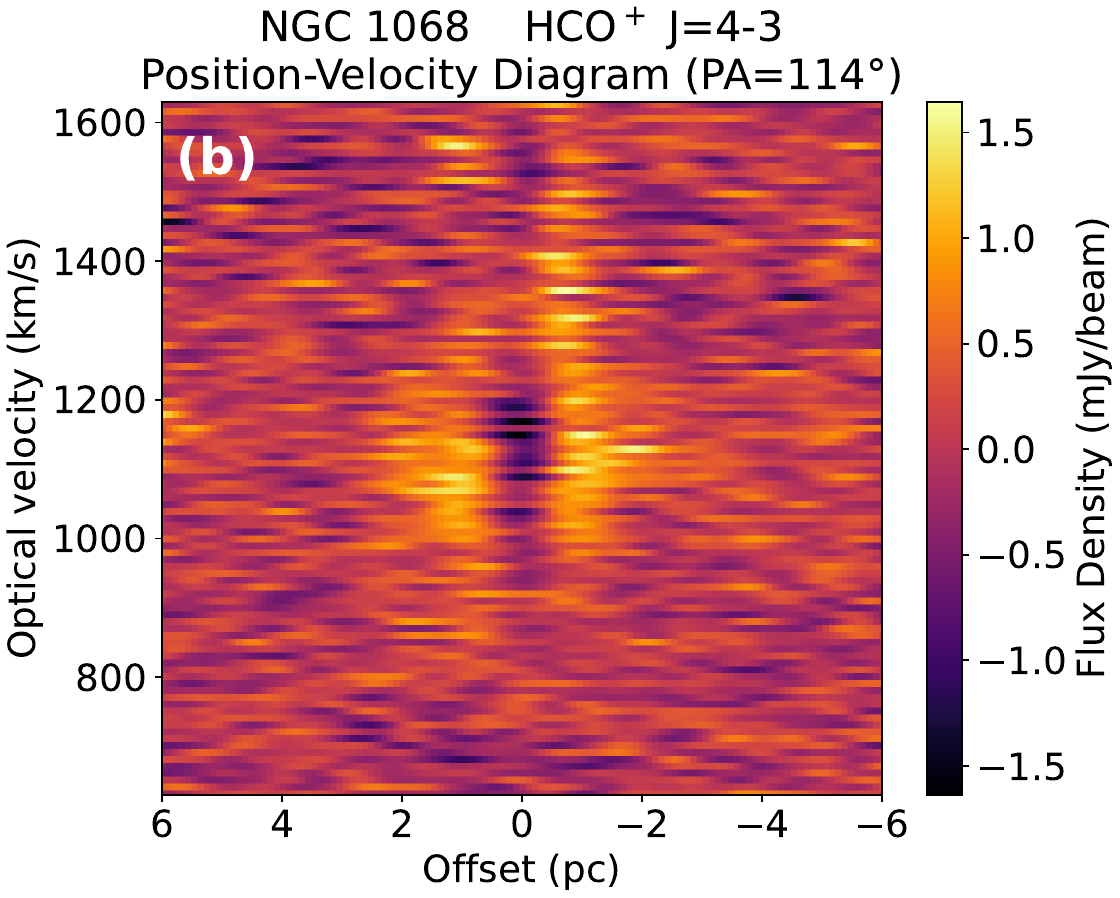} \\
\vspace{0.8cm}
\includegraphics[scale=0.44]{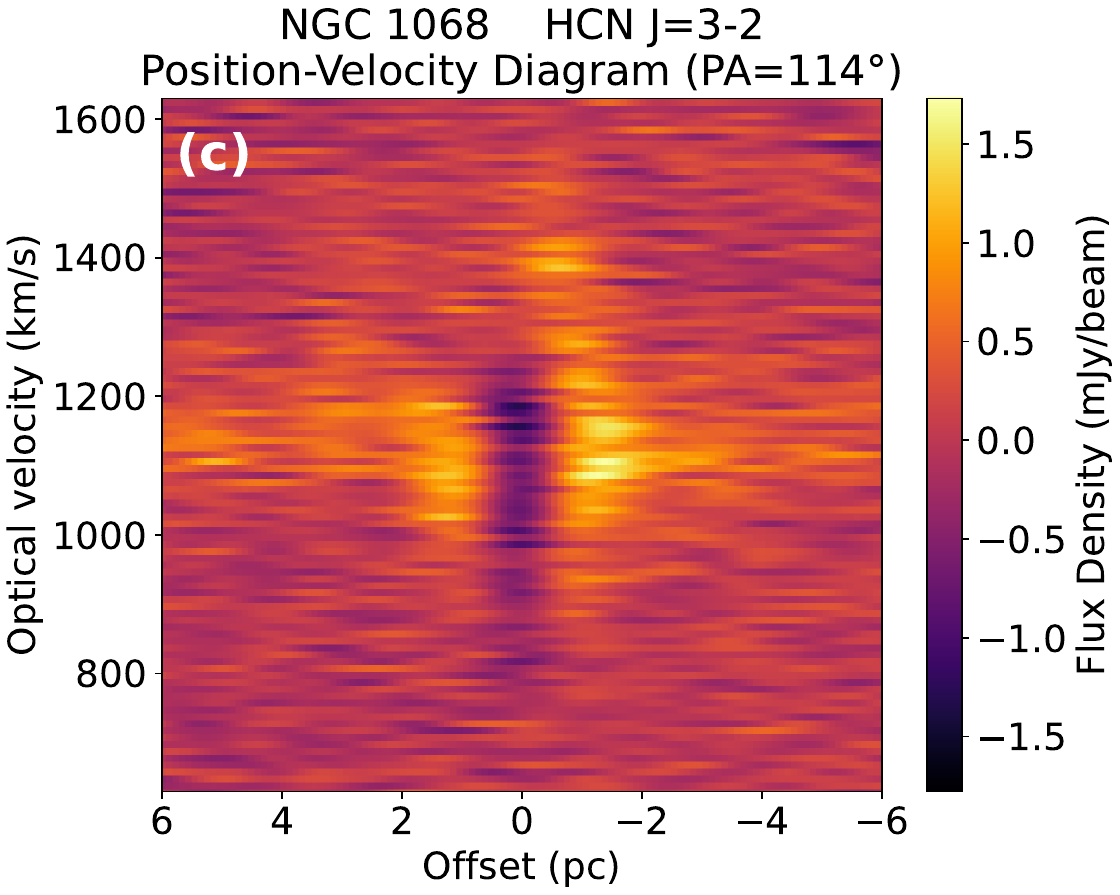} 
\hspace{1.0cm}
\includegraphics[scale=0.44]{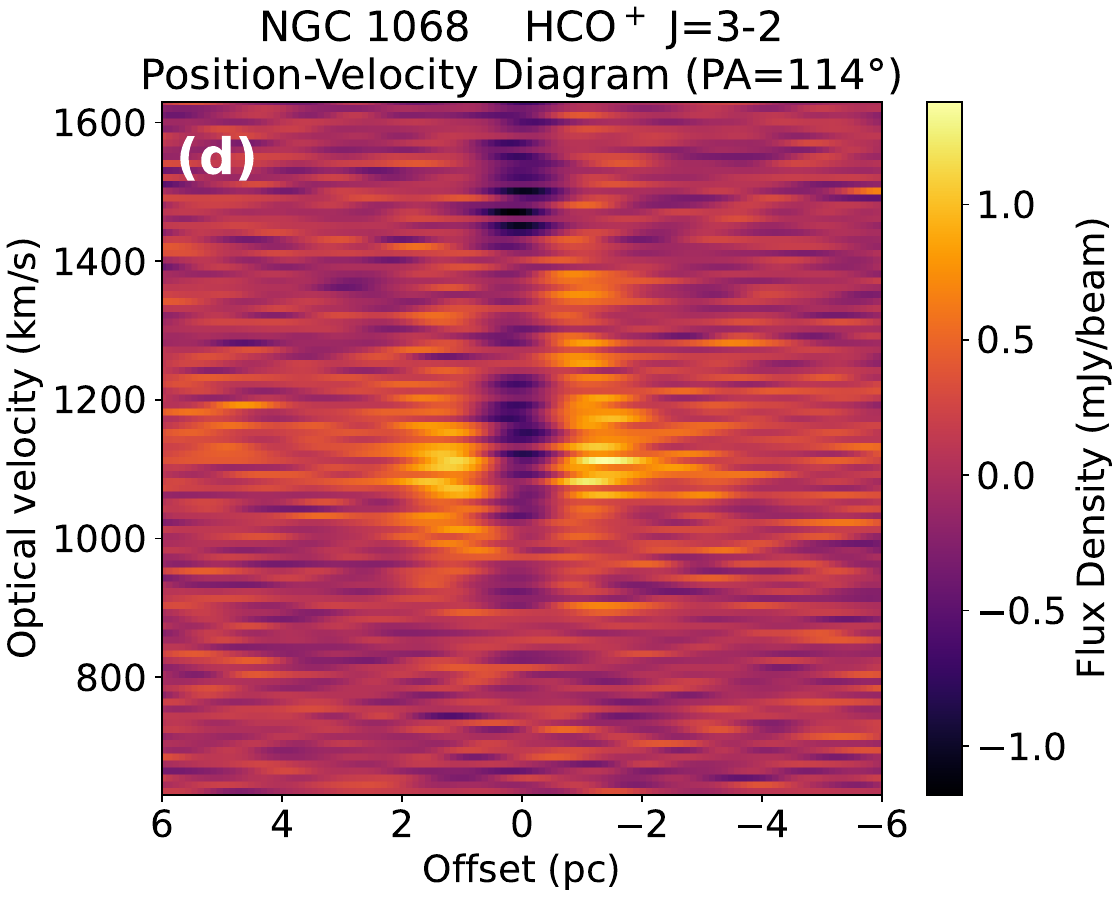} 
\end{center}
\vspace{-0.2cm}
\caption{
Position-velocity diagram of (a) HCN J=4--3,
(b) HCO$^{+}$ J=4--3, (c) HCN J=3--2, and
(d) HCO$^{+}$ J=3--2 lines, along the position angle of
PA = 114$^{\circ}$ east of north.
The abscissa is the offset from the $\sim$260--350 GHz continuum
emission peak (in pc).
West is to the right and east is to the left.
The ordinate is the optical LSR velocity (in km s$^{-1}$).
\label{fig:PV}
}
\end{figure*}

Figure \ref{fig:PV} displays the position-velocity diagrams of the HCN
J=4--3 and HCO$^{+}$ J=4--3 lines, as well as the reanalyzed HCN
J=3--2 and HCO$^{+}$ J=3--2 lines, along the position angle 
PA = 114$^{\circ}$ east of north, following \citet{gam25}.
Significant emission is seen in the redshifted component at V $\sim$
1330--1530 km s$^{-1}$ ($\Delta$V$_{\rm red}$ = $+$200--400 km
s$^{-1}$) in the western torus, as previously found
\citep[e.g.,][]{ima20,vol22,gam25}. 
Absorption components at V $\sim$ 1430--1580 km s$^{-1}$
($\Delta$V$_{\rm red}$ = $+$300--450 km s$^{-1}$) toward the continuum
emission peak (= SMBH position) are visible for the HCO$^{+}$ J=4--3
and HCO$^{+}$ J=3--2 lines (Figure \ref{fig:PV}b,d), but not for the
HCN lines at J=4--3 or J=3--2 (Figure \ref{fig:PV}a,c).

\begin{figure*}[!hbt]
\begin{center}
\hspace{-0.4cm}
\includegraphics[scale=0.37]{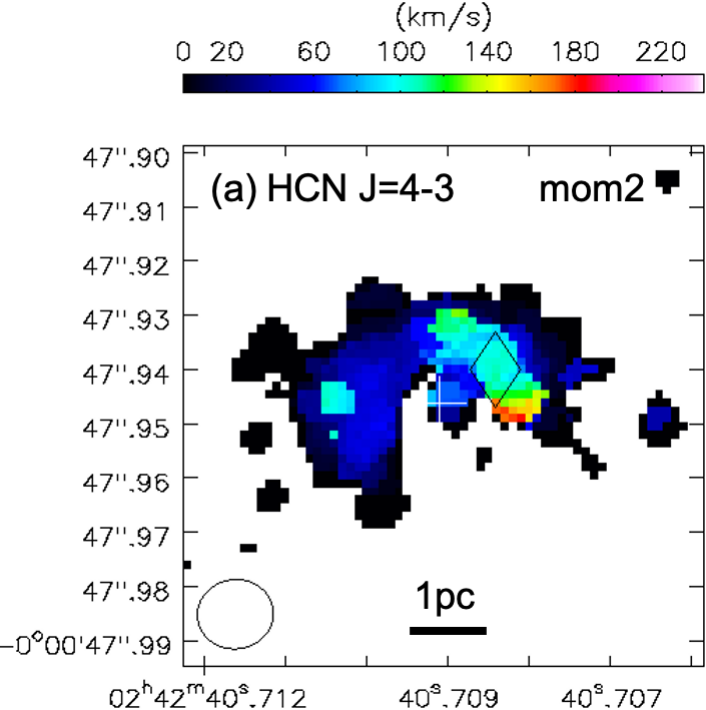} 
\includegraphics[scale=0.37]{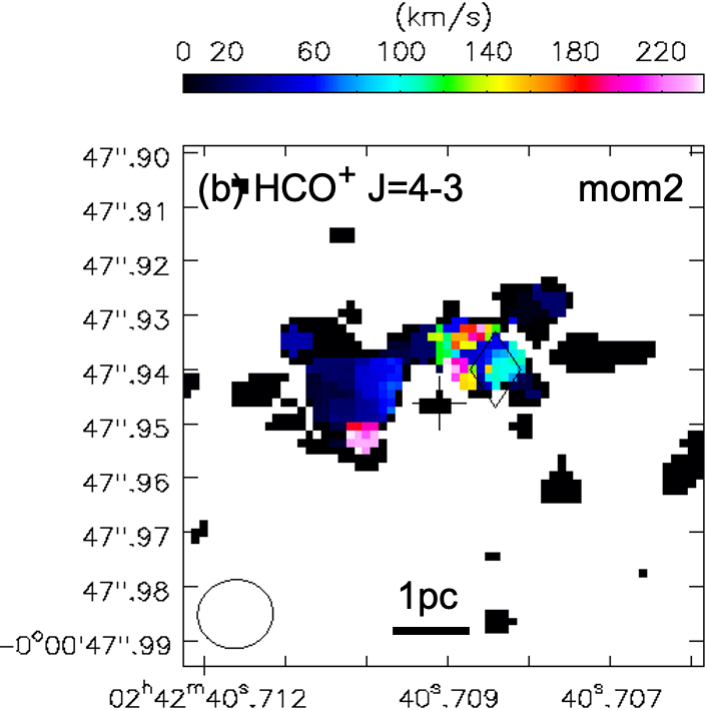} 
\includegraphics[scale=0.37]{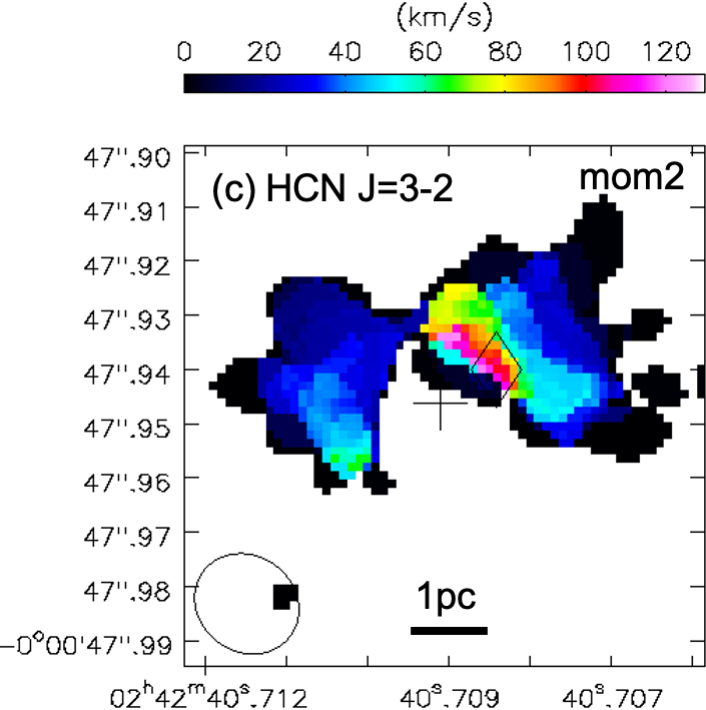} 
\includegraphics[scale=0.37]{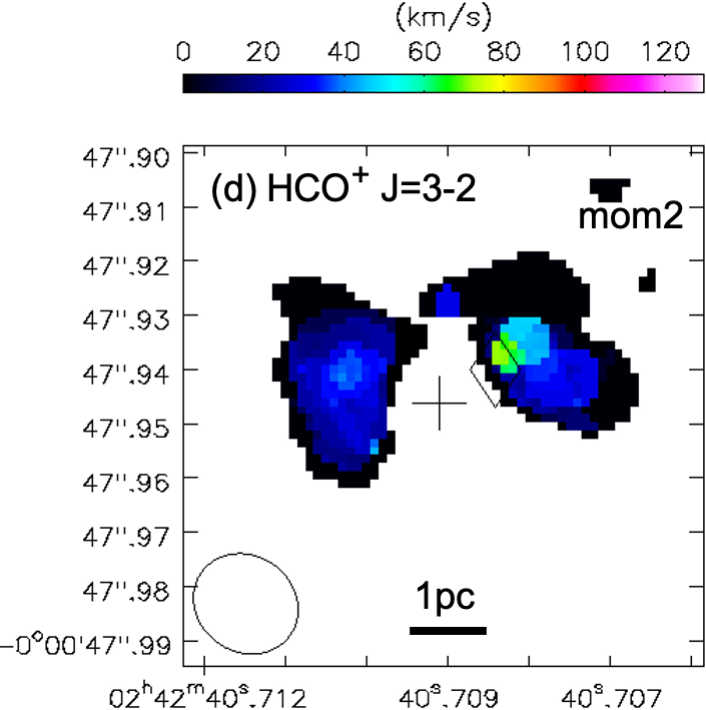} 
\end{center}
\vspace{-0.2cm}
\caption{
Intensity-weighted velocity dispersion (moment 2) map of
(a) HCN J=4--3, (b) HCO$^{+}$ J=4--3, (c) HCN J=3--2, and
(d) HCO$^{+}$ J=4--3, created with 1.5 mas pixel$^{-1}$.
The displayed velocity range is 0--240 km s$^{-1}$ for J=4--3 and
0--130 km s$^{-1}$ for J=3--2.
Note that the peak velocity dispersion value is significantly higher
for the J=4--3 lines than for the J=3--2 lines, possibly because of
the smaller synthesized beam size for J=4--3, which can probe velocity 
structures at a finer spatial resolution.
The horizontal thick bar, open circle, and plus and open diamond
marks are the same as in Figure \ref{fig:mom1}.
\label{fig:mom2}
}
\end{figure*}

Intensity-weighted velocity dispersion (moment 2) maps of the
J=4--3 lines of HCN and HCO$^{+}$, together with the reanalyzed J=3--2 
lines of HCN and HCO$^{+}$, are presented in Figure \ref{fig:mom2}.
The J=4--3 data for HCN and HCO$^{+}$ clearly reveal high velocity
dispersion regions  in the western torus compared to the eastern
torus, as previously found in the J=3--2 data for HCN and HCO$^{+}$
\citep{ima20}. 
The HCN J=4--3 emission peak ``W-peak'' spatially coincides with
the high velocity dispersion regions in the western torus (Figure
\ref{fig:mom2}a--d), as previously seen for HCN J=3--2
\citep{ima20}.

\begin{figure*}[!hbt]
\begin{center}
\hspace{-0.4cm}
\includegraphics[scale=0.37]{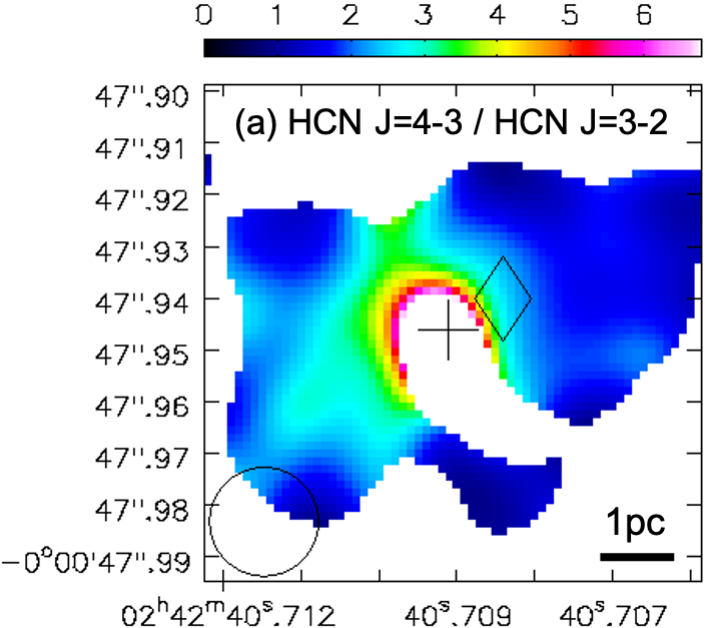} 
\includegraphics[scale=0.37]{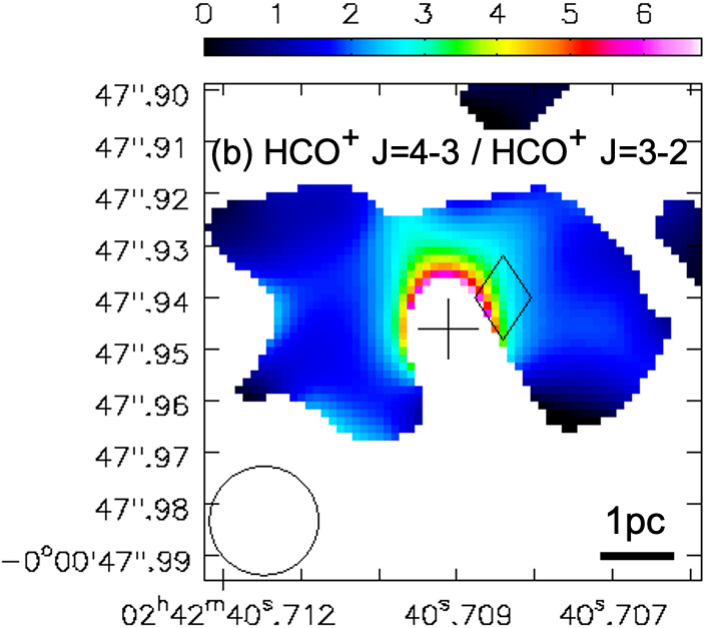} 
\includegraphics[scale=0.37]{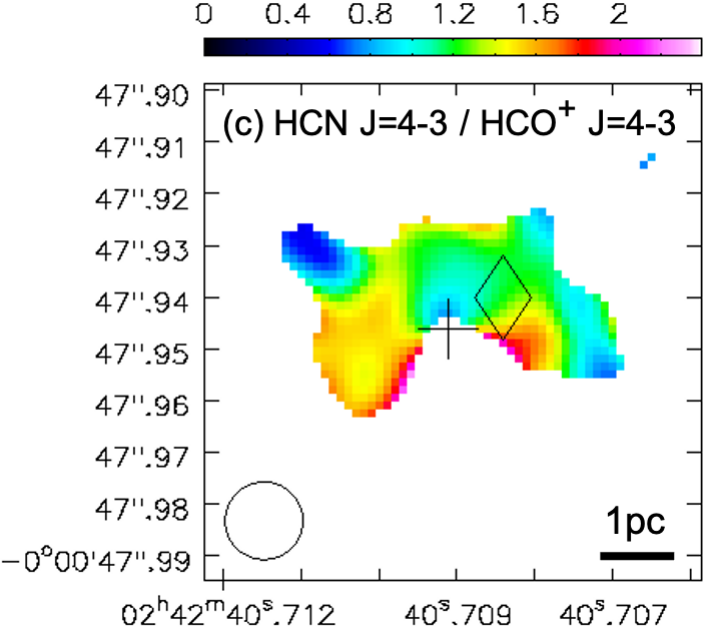} 
\includegraphics[scale=0.37]{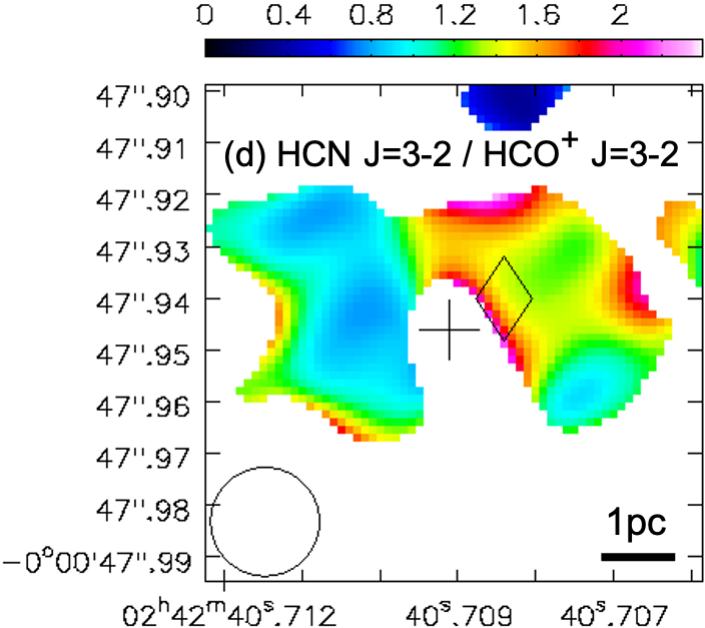} \\
\end{center}
\vspace{-0.2cm}
\caption{
Map of flux ratios calculated in units of Jy beam$^{-1}$ km
s$^{-1}$ for
(a) HCN J=4--3 to HCN J=3--2,
(b) HCO$^{+}$ J=4--3 to HCO$^{+}$ J=3--2,
(c) HCN J=4--3 to HCO$^{+}$ J=4--3, and
(d) HCN J=3--2 to HCO$^{+}$ J=3--2.
Flux ratios are calculated after modifying the synthesized beam to a
0$\farcs$021 circular one for (a), (b), and (d), and a 0$\farcs$015
circular one for (c).
The plus and open diamond marks are the same as in Figure \ref{fig:mom1}.
\label{fig:div}
}
\end{figure*}

Figure \ref{fig:div}a and \ref{fig:div}b present the maps of (a) the HCN
J=4--3 to J=3--2 and (b) the HCO$^{+}$ J=4--3 to J=3--2 flux ratios, 
respectively, after making their synthesized beams identical. 
The flux ratios tend to be higher at the inner part of the torus
than at the outer part, confirming the argument in $\S$3 (paragraph 1)
that the bulk of the J=4--3 emission comes from more compact regions
than the J=3--2 emission for both HCN and HCO$^{+}$.
If the emission is thermalized and optically thick, the expected flux
ratio is 16/9 (=1.8) when calculated in units of Jy beam$^{-1}$ km s$^{-1}$.  
The innermost ($\lesssim$1 pc) regions of both the western and
eastern tori and regions close to the W-peak show flux ratios with
$\gtrsim$2.5. 
Such high values cannot be explained solely by the possible absolute
flux calibration uncertainty of individual ALMA observations
($\lesssim$10\%), which could produce the ratio with maximum $\sim$2.2
(=1.1/0.9 $\times$ 1.8) for the thermalized, optically thick emission.
This suggests that HCN and HCO$^{+}$ in these regions are highly
excited to J=4 and the emission is optically thin (and/or affected by
population inversion), likely influenced by the central AGN. 
The (c) HCN J=4--3 to HCO$^{+}$ J=4--3 and 
(d) HCN J=3--2 to HCO$^{+}$ J=3--2 flux ratios, also calculated in
units of Jy beam$^{-1}$ km s$^{-1}$, are displayed in Figures 
\ref{fig:div}c and \ref{fig:div}d, respectively.
HCN-to-HCO$^{+}$ flux ratios greater than unity at both J=4--3 and
J=3--2 are confirmed across most of the compact ($\lesssim$3--5
pc) region of the NGC 1068 torus in a spatially resolved manner.

\section{Discussion}

\subsection{Summary of multiple scenarios to explain the observed 
dense molecular gas properties in the NGC 1068 torus} 

As briefly described in $\S$1, multiple scenarios have been proposed to
account for the apparent counter-rotation in the compact
($\lesssim$3--5 pc) NGC 1068 torus.
The first is that the inner ($\lesssim$2 pc) and outer ($\gtrsim$2 pc)
parts of the NGC 1068 torus are indeed counter-rotating with respect to
each other \citep[e.g.,][]{imp19,ima20}, possibly because of a past
collision with a very compact gas clump that had sufficient angular
momentum and approached the western side of the originally rotating
torus from the opposite direction \citep[e.g.,][]{ima20,vol22}.
This configuration is dynamically unstable and short-lived
($<$10$^{5}$ yr) \citep{vol22}. 
Thus, the probability of detecting this physically counter-rotating
phase is very small \citep[e.g.,][]{gar19}, but it may still be
detectable \citep{vol22}.
In this physically counter-rotating gas scenario, angular momentum of
the gas in the torus can be efficiently removed, enabling a high mass
accretion rate onto the central SMBH
\citep[e.g.,][]{kuz99,qua15,dyd15,vor15}, naturally explaining the
fact that NGC 1068 is observed as a luminous AGN 
(L$_{\rm AGN}$ $\gtrsim$ 10$^{45}$ erg s$^{-1}$)
\citep[e.g.,][]{boc00,hon08,rab09}. 

The second scenario is that the observed apparent counter-rotation
signatures could be produced by outflow activity
\citep[e.g.,][]{gar19,wil20,ban23}.
This outflow-origin scenario is longer-lived and therefore more likely
to be observed than the first physically counter-rotating scenario in
terms of timescale.
However, it remains unclear what fundamental physical mechanism allows
NGC 1068 to sustain its luminous AGN activity by removing the angular
momentum of gas in the torus and enabling a high mass accretion rate
onto the central SMBH. 
Furthermore, the significant asymmetry of molecular gas
emission observed between the western and eastern tori is difficult to
reproduce solely with this outflow-origin model \citep{gam25}.

\citet{gam25} recently proposed a third scenario involving an infalling
gas cloud from the western side to the immediate vicinity of the
mass-accreting SMBH, without strong interaction with the torus. 
However, these authors focused only on (1) the observed asymmetry of
molecular gas emission between the western and eastern tori, 
and (2) absorption features
detected for the CO J=3--2 and HCO$^{+}$ J=4--3 lines at
$\sim$400 km s$^{-1}$ redshifted relative to the
systemic velocity of NGC 1068 (V$_{\rm sys}$ = 1130 km s$^{-1}$),
toward the $\sim$350 GHz continuum emission peak \citep{gam25}.
The overall properties of the torus itself were not sufficiently
discussed in the context of this third scenario.

We attempt to place further constraints on the dense molecular gas
emission and dynamical properties of the NGC 1068 torus, particularly
the western torus, by using our highest-spatial-resolution
($\lesssim$1 pc) bright HCN J=4--3 line data, combined with the
$\lesssim$1 pc-resolution HCO$^{+}$ J=4--3 and $\sim$1.5 pc-resolution
HCN J=3--2 and HCO$^{+}$ J=3--2 line data.

\subsection{Highly turbulent dense molecular gas at the HCN J=4--3 
emission peak of the torus}

We estimate the emission line luminosities of HCN J=4--3, HCO$^{+}$
J=4--3, HCN J=3--2, and HCO$^{+}$ J=3--2 in the western torus
(1.7 pc diameter around the W-peak), the eastern torus (1.7 pc diameter
around the E-peak), and the combined torus region (6.8 pc diameter
around the C-peak) from their fluxes (Table \ref{tab:Gauss}), using
the following equation \citep{sol05},
\begin{eqnarray}
\left(\frac{L'_{line}}{\rm K\ km\ s^{-1}\ pc^{2}}\right) =
3.25 \times 10^{7}
\left(\frac{\nu_{rest}}{\rm GHz}\right)^{-2} (1+z)^{-1}
\left(\frac{D_{L}}{\rm Mpc}\right)^2
\left(\frac{S \Delta V}{\rm Jy\ km\ s^{-1}}\right),
\end{eqnarray}
where S$\Delta$V is the Gaussian-fit, velocity-integrated emission
line flux (in Jy km s$^{-1}$) and D$_{L}$ is the luminosity distance
(in Mpc).
Table \ref{tab:luminosity} summarizes the derived molecular emission
line luminosities.

\begin{deluxetable*}{ccccc}[!hbt]
\tabletypesize{\scriptsize}
\tablecaption{Molecular Emission Line Luminosity \label{tab:luminosity}} 
\tablewidth{0pt}
\tablehead{
\colhead{Torus position} & \colhead{HCN J=4--3} & \colhead{HCO$^{+}$ J=4--3} & 
\colhead{HCN J=3--2} & \colhead{HCO$^{+}$ J=3--2} \\
\colhead{} & \colhead{10$^{4}$ (K km s$^{-1}$ pc$^{2}$)} & 
\colhead{10$^{4}$ (K km s$^{-1}$ pc$^{2}$)} & 
\colhead{10$^{4}$ (K km s$^{-1}$ pc$^{2}$)} &  
\colhead{10$^{4}$ (K km s$^{-1}$ pc$^{2}$)} \\
\colhead{(1)} & \colhead{(2)} & \colhead{(3)}  & \colhead{(4)}  &
\colhead{(5)} 
}
\startdata 
W-torus (1.7 pc diameter) & 4.8$\pm$0.5 & 3.0$\pm$0.5 &
2.8$\pm$0.6 & 1.6$\pm$0.2 \\  
E-torus (1.7 pc diameter) & 3.7$\pm$0.3 & 2.3$\pm$0.3 &
1.5$\pm$0.2 & 1.2$\pm$0.2 \\
Torus (6.8 pc diameter) & 21.7$\pm$1.5 & 12.5$\pm$1.0 & 18.0$\pm$0.9 &
12.4$\pm$0.9 \\
\enddata

\tablecomments{
Col.(1): Torus region.
Cols.(2)--(5): Emission line luminosities of
(2) HCN J=4--3, (3) HCO$^{+}$ J=4--3, (4) HCN J=3--2, and (5)
HCO$^{+}$ J=3--2, in units of 10$^{4}$ K km s$^{-1}$ pc$^{2}$.
}

\end{deluxetable*}

Because the critical densities of the HCN and HCO$^{+}$ lines at
J=4--3 and J=3--2 are n$_{\rm crit}$ $\gtrsim$ 10$^{6}$ cm$^{-3}$ at a
gas kinetic temperature of 50--100 K \citep{shi15}, only dense and
warm molecular gas can sufficiently excite HCN and HCO$^{+}$ to J=4
and J=3. 
For most molecular gas in luminous starburst galaxies and AGNs,
HCN and HCO$^{+}$ luminosities in units of K km s$^{-1}$ pc$^{2}$ at J=4--3
are usually smaller than those at J=3--2 if these molecules are
sub-thermally excited and the emission is optically thick
\citep[e.g.,][]{knu07,kri08,gre09,pap14,isr23,ima23a,ima23b}.
The luminosities in these units can be comparable between J=4--3 and
J=3--2
if HCN and HCO$^{+}$ are thermally excited and the emission is optically
thick.
In Table \ref{tab:luminosity}, the J=4--3 and J=3--2 luminosities
in the 6.8 pc diameter circular region centered on the mass-accreting
SMBH position are roughly comparable, within $\lesssim$20\%,
suggesting that HCN and HCO$^{+}$ are nearly thermally excited to J=4
and J=3, and that the emission is optically thick when integrating molecular
gas emission over the entire NGC 1068 compact ($\lesssim$3--5 pc) torus
region.

However, in both the western and eastern tori with 1.7 pc diameter
circular regions around the HCN J=4--3 emission peaks, the luminosities of
both HCN and HCO$^{+}$ are a factor of $\gtrsim$1.7 higher 
\footnote{
The possible absolute flux calibration uncertainty of individual ALMA 
observations ($\lesssim$10\%) again could increase the J=4--3 to
J=3--2 flux ratio by a factor of only $\lesssim$1.2 (= 1.1/0.9).
}
at J=4--3 than at J=3--2 (Table \ref{tab:luminosity}),
suggesting that HCN and HCO$^{+}$ are nearly thermally excited and the
emission is optically thin in the small areas around the W-peak and
E-peak. 
Large velocity dispersion caused by high turbulence can reduce
molecular line opacity and reproduce the observed HCN and
HCO$^{+}$ emission line properties in these regions.
The first physically counter-rotating torus scenario, caused by
a physical collision of a compact dense gas clump with the
western side of the rotating dense molecular torus from the opposite
direction \citep[e.g.,][]{ima20,vol22}, can naturally generate high
turbulence in the torus, particularly in small regions around the
collision interface, originating from the Kelvin-Helmholtz
instabilities \citep[e.g.][]{qua15,dyd15}. 
The spatial coincidence between the observed large velocity dispersion
regions and the W-peak (Figure \ref{fig:mom2}a--d) can be explained if
the putative collision interface in the western torus is close to the
W-peak, where the observed velocity dispersion values can increase
through the combination of (i) enhancement of intrinsic velocity
dispersion produced by the turbulence and (ii) beam smearing of
two counter-rotating gas components. 
However, the third scenario of an infalling gas cloud from the west
to the SMBH vicinity, without physical collision with the torus
\citep{gam25}, has no direct way to reduce the opacity of molecular
line emission in the small areas around the W-peak (and E-peak), which
is suggested by our new observations. 

The physical connection between the redshifted 22 GHz H$_{2}$O maser
and the HCN J=4--3 dense molecular line emission at the innermost western
torus, suggested by their comparable velocity profiles
(Figure \ref{fig:Wtorus}), also supports the first physically
counter-rotating dense molecular gas scenario (particularly in the
western torus) \citep{ima20}, but cannot be naturally explained by 
the third scenario, in which the redshifted molecular line emission from
the innermost western torus is produced by an infalling gas cloud
physically unrelated to the H$_{2}$O maser-emitting rotating disk
\citep{gam25}, as well as outflow-origin models that do not
invoke a rotating dense molecular torus \citep{wil20,ban23}.  
We thus further explore the first collision-induced physically
counter-rotating torus molecular gas scenario.

\subsection{Comparison of the physically counter-rotating torus  
scenario of \citet{vol22} with the highest-spatial-resolution dense
molecular gas data} 

In the collision-induced physically counter-rotating torus scenario,
two more observational characteristics may be produced at the
interface between the two counter-rotating gas components, 
particularly in the western torus, in addition to the observed
velocity dispersion increase ($\S$4.2).
First, a possible 0.5--1 pc wide gap may be created
\citep[e.g.,][]{vol22}.
However, we see no clear signature of such a gap even in
our highest-spatial-resolution ($\lesssim$1 pc) integrated-intensity
(moment 0) map of the bright HCN J=4--3 emission
(Figure \ref{fig:mom0}a).
Second, excitation of dense molecular gas can be enhanced by 
turbulence-induced shocks.
We do not see any clear sign of elevated J=4--3 to J=3--2 flux ratios
of HCN and HCO$^{+}$ emission, driven by the putative shocks, at the
W-peak (Figure \ref{fig:div}).

In summary, even using our new $\lesssim$1 pc resolution bright HCN
J=4--3 line data, we cannot provide definitive observational evidence
to conclusively support the first collision-induced physically
counter-rotating torus scenario, possibly because the achieved
$\lesssim$1 pc resolution is still insufficient to investigate the
predicted collision interface between two counter-rotating gas
components in detail.
Even higher-spatial-resolution ($\ll$1 pc) data may help better
scrutinize the putative interface regions.

Even though we cannot clearly detect the putative collision interface,
our new $\lesssim$1 pc ($\lesssim$14 mas)-resolution HCN J=4--3 and
HCO$^{+}$ J=4--3 line data can be used to update the collision-induced
physically counter-rotating torus model of \citet{vol22}.
\citet{vol22} compared the predictions of their analytical model with
the ALMA molecular line data available at that time, including multiple CO
J-transition line data and $\sim$1.5 pc ($\sim$20 mas) resolution HCN
J=3--2 and HCO$^{+}$ J=3--2 line data, and concluded that 
the observed overall molecular line properties can be quantitatively
reproduced by the model within a factor of $\sim$2. 
However, only low-spatial-resolution HCN J=4--3 ($\gtrsim$7 pc 
or $\gtrsim$100 mas) and HCO$^{+}$ J=4--3 ($\gtrsim$2 pc or $\gtrsim$30
mas) data were used.
For HCN J=4--3, no detailed comparison was possible.
For HCO$^{+}$ J=4--3, an absorption feature toward the continuum
emission peak at the systemic velocity (V$_{\rm sys}$ = 1130 km
s$^{-1}$) was predicted by the model but was not seen
in the observed data (Fig. 20 of \citet{vol22}), most likely because
it was diluted by emission in the low-spatial-resolution data.
We now have new higher-spatial-resolution ($\lesssim$1 pc or 14 mas)
HCN J=4--3 and HCO$^{+}$ J=4--3 line data, enabling an updated
comparison with the model.
Indeed, the predicted absorption feature at the systemic velocity
(V$_{\rm sys}$ = 1130 km s$^{-1}$) toward the continuum emission
peak was detected in our beam-sized spectra
(Figure \ref{fig:spectraPeakVopt}) and position velocity diagrams
(Figure \ref{fig:PV}) of the HCN J=4--3 and HCO$^{+}$ J=4--3 lines.

\begin{figure*}[!hbt]
\begin{center}
\includegraphics[scale=0.124]{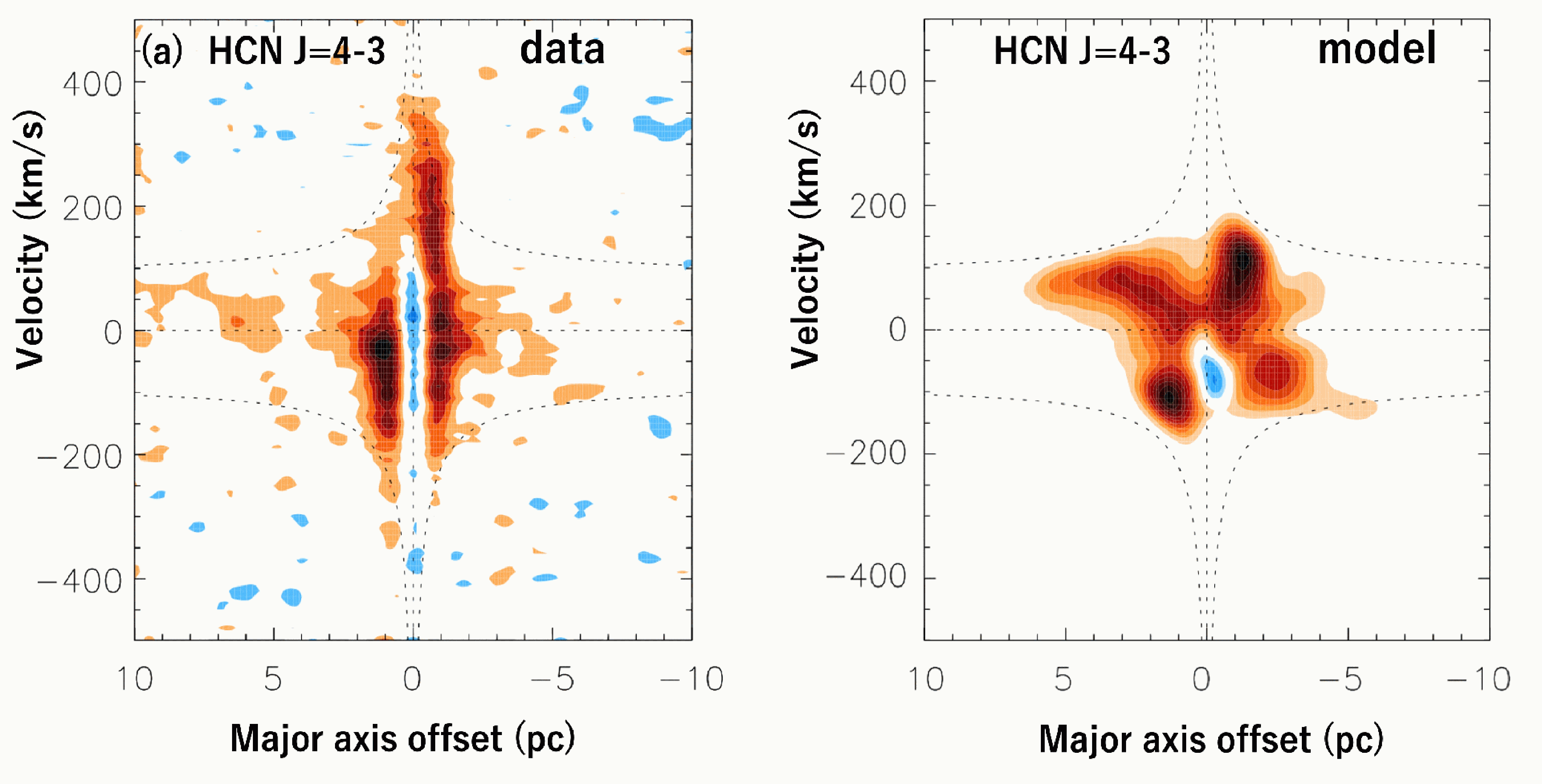}  
\hspace{0.4cm}
\includegraphics[scale=0.124]{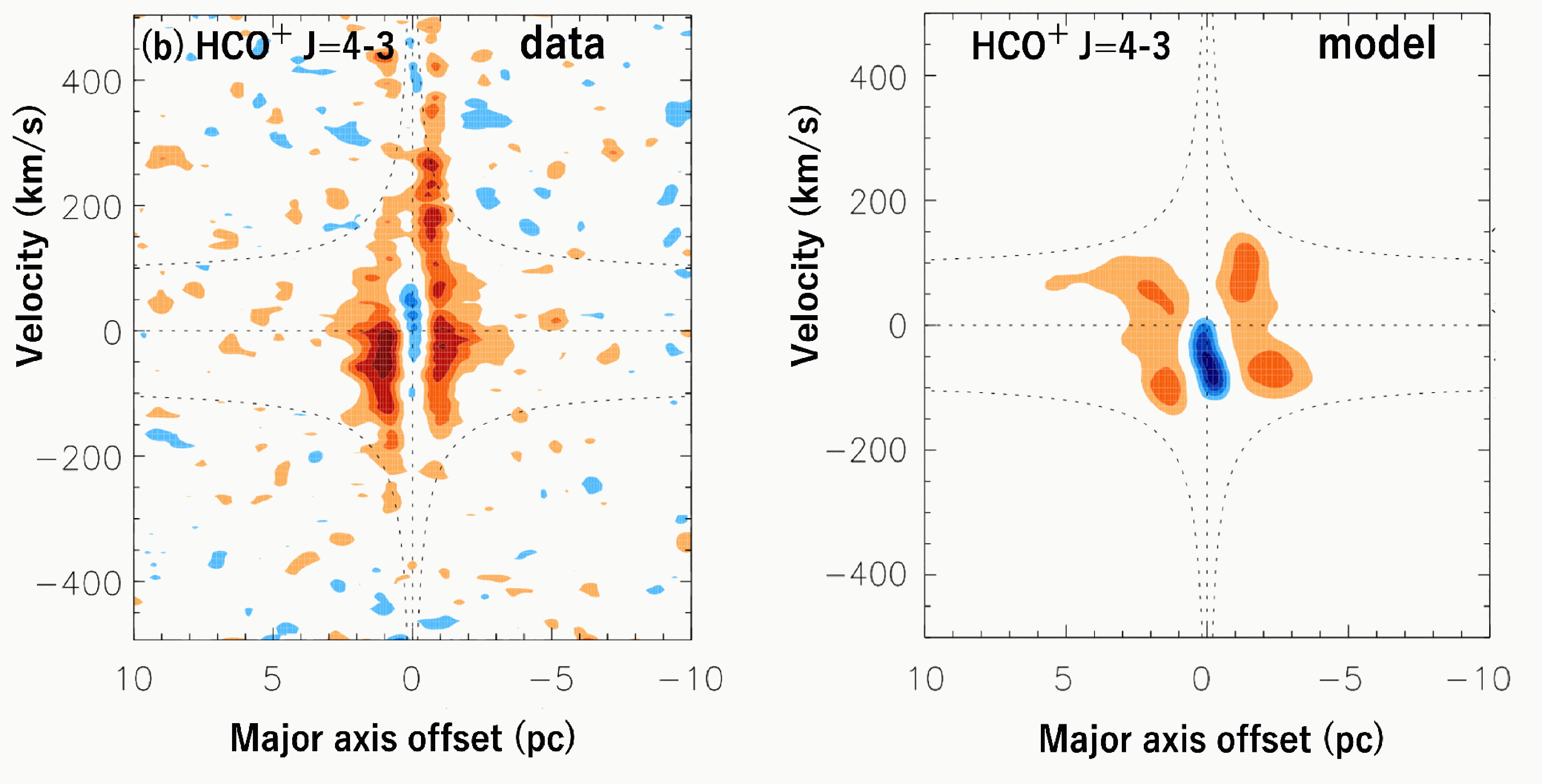} 
\end{center}
\vspace{-0.2cm}
\caption{
Comparison of position-velocity diagrams between observational data and
the model by \citet{vol22}.
{\it (a)}: HCN J=4--3. {\it (b)}: HCO$^{+}$ J=4--3.
In these plots, the major axis is PA = 110$^{\circ}$, consistent with
those shown in \citet{vol22}. 
For the observations {\it (Left)}, the images are rotated by
20$^{\circ}$ to make the elongated structure horizontal (E-W).
For the model {\it (Right)}, the data are projected such that the disk
plane is horizontal (E-W).
In each panel, the curved dotted line indicates the expected Keplerian
rotation for the radial distribution of the total enclosed mass,
M(R) = M$_{\rm SMBH}$ $+$ M$_{0}$R$^{5/4}$, where M$_{\rm SMBH}$ is the
central SMBH mass (M$_{\rm SMBH}$ = 1 $\times$ 10$^{7}$M$_{\odot}$) 
and M$_{0}$ is the stellar mass distribution 
(M$_{0}$ = 0.8 $\times$ 10$^{6}$M$_{\odot}$ [pc$^{-5/4}$]), in the
same way as presented in \citet{vol22}.
\label{fig:VollmerPV}
}
\end{figure*}

\begin{figure*}[!hbt]
\begin{center}
\includegraphics[scale=0.27]{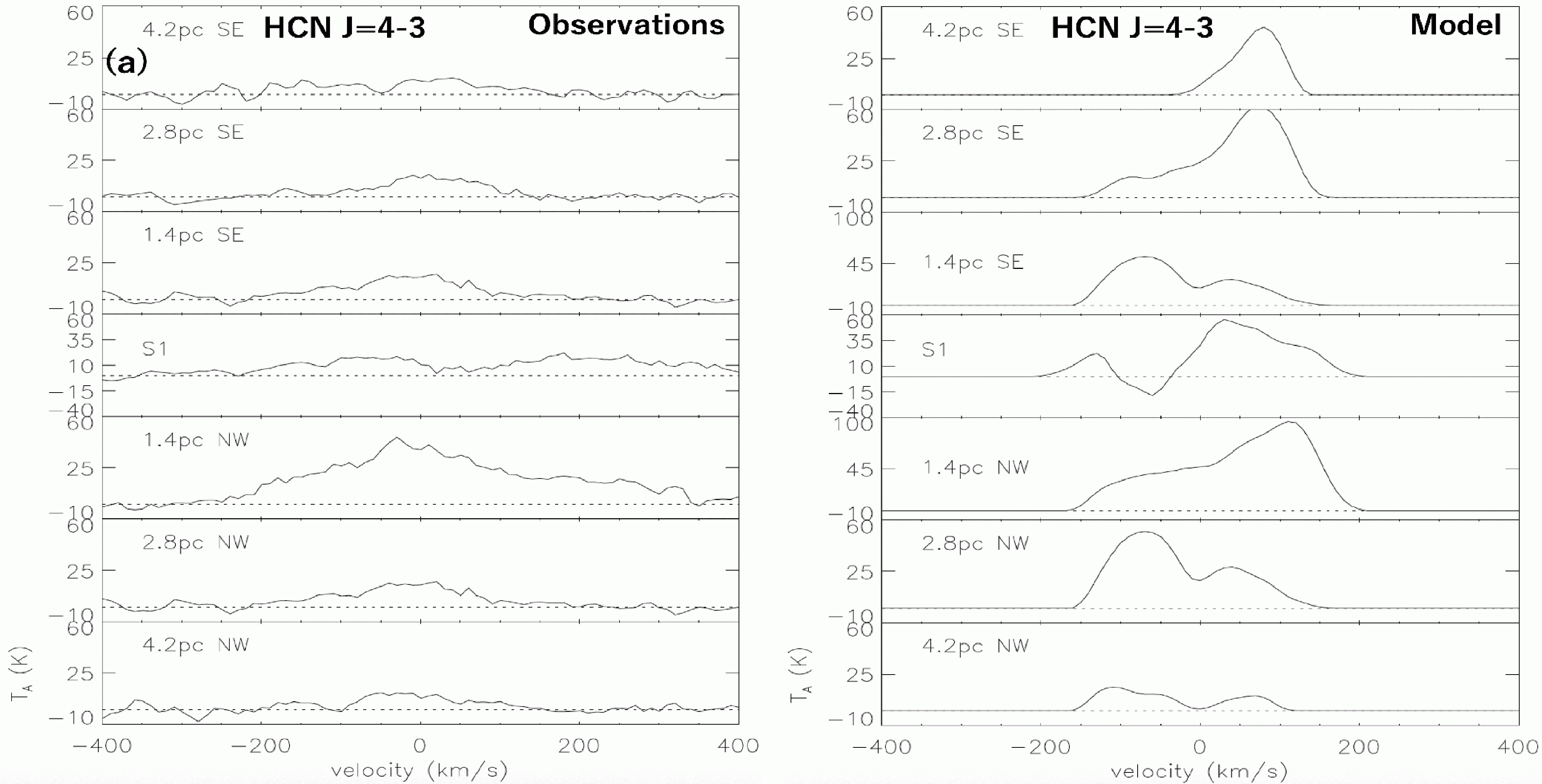}  \\
\vspace*{1.0cm}
\includegraphics[scale=0.27]{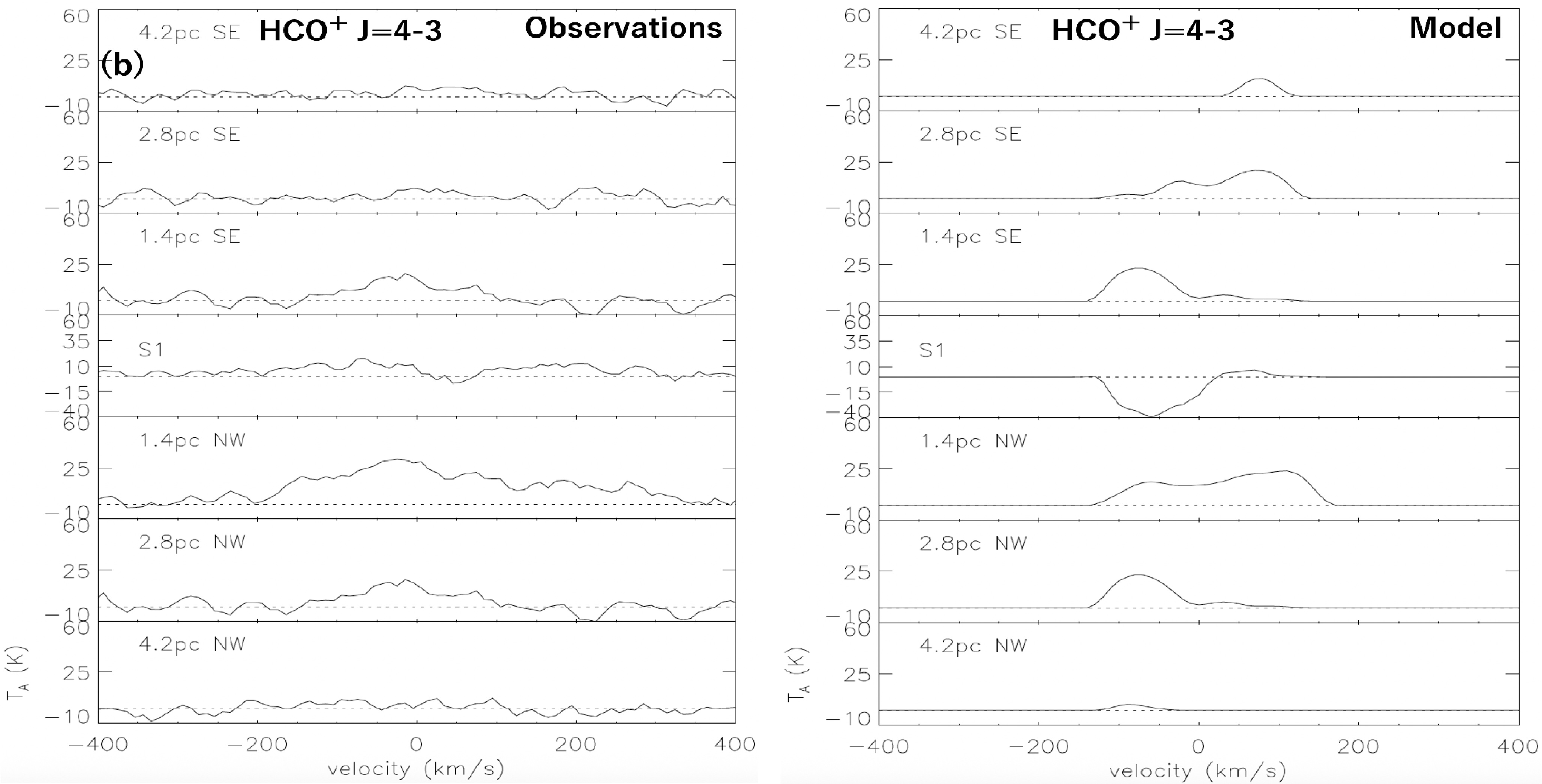} 
\end{center}
\caption{
Comparison between observational {\it (Left)} and model {\it (Right)}
spectra by \citet{vol22}, at multiple torus positions.
{\it (a)}: HCN J=4--3. {\it (b)}: HCO$^{+}$ J=4--3.
In these plots, NW and SE are along PA = 110$^{\circ}$, to be
consistent with those shown in \citet{vol22}.
Spectra are extracted along the NW and SE directions, at distances of
0 pc, $\pm$1.4 pc, $\pm$2.8 pc, and $\pm$4.2 pc, with 1.4 pc diameter
circular apertures.
\label{fig:VollmerSpectra}
}
\end{figure*}

Figure \ref{fig:VollmerPV} shows a detailed comparison of the 
position-velocity diagrams of the HCN J=4--3 and HCO$^{+}$ J=4--3
lines between our new $\lesssim$1 pc resolution data and the best-fit
model of \citet{vol22}. 
The main features of the model prediction (i.e., absorption at the
systemic velocity) are confirmed by our new $\lesssim$1 pc resolution
HCN J=4--3 and HCO$^{+}$ J=4--3 line data.
Observed and predicted spectra at multiple positions along the major
axis (PA = 110$^{\circ}$) are compared in Figure
\ref{fig:VollmerSpectra}.
The overall properties largely agree between the model and
observations, as in the case for other molecular lines
\citet{vol22}. 
The observed east-west asymmetry of dense molecular line emission
appears not to be adequately reproduced by the best-fit model of
\citet{vol22}, but this discrepancy largely comes from initial
conditions of the modeling. 
Further model refinement is important, because this collision model,
consisting of infalling gas clump to the western torus, should be able
to create significant east-west asymmetry, which is very difficult to
naturally explain by the outflow-induced model   
\citep[e.g.,][]{gar19,wil20,ban23}.

\subsection{Further confirmation of enhanced HCN abundance in the 
compact nuclear region of NGC 1068}

\subsubsection{From nuclear outflow activity}

Nuclear dense molecular outflow activity has been proposed
based on the detection of a blueshifted HCN J=3--2 absorption wing 
with up to $\Delta$$V_{\rm blue}$ $\sim$ $-$400 km s$^{-1}$ in the
spectrum toward the $\sim$260 GHz continuum emission peak (the putative
location of the mass-accreting SMBH) \citep{imp19,ima20}. 
This blueshifted absorption wing was less clear for HCO$^{+}$
J=3--2 \citep{ima20} and was reported to be undetected for HCO$^{+}$
J=4--3 \citep{gam25}.
The blueshifted absorption wing reaching $\Delta$$V_{\rm blue}$
$\sim$ $-$400 km s$^{-1}$ (V $\sim$ 730 km s$^{-1}$) is recognizable
for HCN J=4--3 (Figure \ref{fig:spectraPeakVopt}a) but appears weaker
for HCO$^{+}$ J=4--3 (Figure \ref{fig:spectraPeakVopt}b).
The stronger observed outflow-origin absorption wings for HCN than
those for HCO$^{+}$ can be explained by the enhanced HCN abundance
relative to HCO$^{+}$ found in the compact ($\lesssim$3--5 pc) nuclear
region around the luminous AGN in NGC 1068
\citep[e.g.,][]{ima18a,ima20,vol22,but22}.

\subsubsection{From torus dense molecular H$_{2}$ gas mass estimate}

We estimate the HCN J=4--3 emission line flux within the 6.8 pc
diameter aperture around the central continuum-emitting AGN
to be 4.3$\pm$0.3 Jy km s$^{-1}$ (Table \ref{tab:Gauss}).
This recovers $\gtrsim$80\% of the integrated HCN J=4--3 flux from the
torus (5.3 Jy km s$^{-1}$) reported in Table 4 of \citet{vol22}.
The HCO$^{+}$ J=4--3 emission line flux within the same 6.8 pc
aperture is estimated to be 2.5$\pm$0.2 Jy km
s$^{-1}$ (Table \ref{tab:Gauss}), again recovering $\gtrsim$80\% of
the integrated HCO$^{+}$ J=4--3 flux (3.0 Jy km s$^{-1}$) in
Table 4 of \citet{vol22}.
Thus, most of the J=4--3 emission of HCN and HCO$^{+}$ from the entire NGC
1068 torus originates in the compact ($\lesssim$3--5 pc)
region.

Dense molecular hydrogen (H$_{2}$) mass is often estimated from
(optically thick) HCN J=1--0 line luminosity using the formula
M$_{\rm dense-H_2}$(HCN) = 6--12 $\times$ L(HCN J=1--0)
(M$_{\odot}$ [K km s$^{-1}$ pc$^{2}$]$^{-1}$)
\citep[e.g.,][]{gao04,kri08,ler17}.
On the compact torus scale (6.8 pc diameter aperture),
J=4--3 and J=3--2 emission line luminosities in units of
(K km s$^{-1}$ pc$^{2}$) are roughly comparable for both HCN and
HCO$^{+}$ (Table \ref{tab:luminosity}),
suggesting that they are thermalized and their emission are optically
thick. 
Assuming that the HCN J=1--0 luminosity in these units is also comparable
to that of HCN J=3--2 and J=4--3 ($\sim$20 $\times$ 10$^{4}$; Table
\ref{tab:luminosity}), 
we obtain HCN-derived torus dense molecular H$_{2}$  gas mass of
M$_{\rm dense-H_2}$(HCN) = (1.2--2.4) $\times$ 10$^{6}$ M$_{\odot}$.

Dense H$_{2}$ mass is also estimated from
HCO$^{+}$ J=1--0 line luminosity using the formula 
M$_{\rm dense-H_2}$(HCO$^{+}$) = 2--5 $\times$ L(HCO$^{+}$ J=1--0)
(M$_{\odot}$ [K km s$^{-1}$ pc$^{2}$]$^{-1}$) in the
optically thick regime \citep[e.g.,][]{ler17}.
Assuming that the HCO$^{+}$ J=1--0 luminosity in these units is comparable
to that of HCO$^{+}$ J=4--3 and J=3--2
($\sim$12 $\times$ 10$^{4}$; Table \ref{tab:luminosity}),
we obtain HCO$^{+}$-derived torus dense H$_{2}$ mass of
M$_{\rm dense-H_2}$(HCO$^{+}$) = (2.4--6.0) $\times$ 10$^{5}$ M$_{\odot}$.
This is a factor of $\sim$4--5 smaller than the HCN-derived dense
H$_{2}$ mass and can be explained by the enhanced HCN abundance
scenario in the compact torus region of NGC 1068
\citep[e.g.,][]{ima18a,ima20,but22}, because standard HCN and HCO$^{+}$
abundances are assumed for the above conversion from line luminosity
to dense H$_{2}$ mass.
The HCO$^{+}$-derived torus dense H$_{2}$ mass is roughly
comparable to the value adopted (3.6 $\times$ 10$^{5}$ M$_{\odot}$) in
the modeling of \citet{vol22} (their Table A.1).

\section{Summary}

We presented the results of our new ALMA $\lesssim$1 pc ($\lesssim$14 mas)
resolution HCN J=4--3 and HCO$^{+}$ J=4--3 line observations of NGC
1068.
The achieved spatial resolution at J=4--3 is $\gtrsim$30\% better than
that at J=3--2, because of its shorter wavelength (higher frequency).
By combining these new $\lesssim$1 pc resolution J=4--3 data with
existing and reanalyzed $\sim$1.5 pc resolution HCN J=3--2 and
HCO$^{+}$ J=3--2 data, we obtained the following main results.

\begin{enumerate}

\item
HCN J=4--3 and HCO$^{+}$ J=4--3 dense molecular line emission is
clearly detected in the compact ($\lesssim$3--5 pc) nuclear region
along the almost east-west direction from the $\sim$350 GHz continuum
emission peak.
We interpret this molecular emission as originating from the putative
dense molecular torus around a mass-accreting SMBH.

\item
In both the western and eastern tori, HCN J=4--3 emission is
brighter than HCO$^{+}$ J=4--3 emission, as previously seen at J=3--2.
The western torus is brighter than the eastern torus in both HCN
J=4--3 and HCO$^{+}$ J=4--3 emission lines.

\item
In the western torus, the inner ($\lesssim$2 pc) part is redshifted
and the outer ($\gtrsim$2 pc) part is blueshifted, with respect to the 
systemic velocity of NGC 1068 (V$_{\rm sys}$ = 1130 km s$^{-1}$), for
both HCN J=4--3 and HCO$^{+}$ J=4--3 emission, confirming apparent
counter-rotation as previously found for HCN J=3--2 and
HCO$^{+}$ J=3--2 lines.

\item
In the eastern torus, both HCN J=4--3 and HCO$^{+}$ J=4--3
emission are blueshifted at the inner ($\lesssim$2 pc) part, as
previously identified at J=3--2 of HCN and HCO$^{+}$.
However, a redshifted emission component at the outer ($\gtrsim$2 pc)
part, previously detected in HCN J=3--2 and CO J=3--2 lines, is
not clearly seen in either HCN J=4--3 or HCO$^{+}$ J=4--3, suggesting
that the outer redshifted molecular gas in the eastern torus is not
sufficiently warm and dense.

\item
The spatial and dynamical properties of the innermost ($\lesssim$1 pc)
redshifted HCN J=4--3 emission in the western torus largely agree with
those of the even inner (0.5--1.2 pc) 22 GHz H$_{2}$O maser emission
spots, previously identified with VLBI very-high-angular-resolution
($\lesssim$2 mas) observations.
This suggests that in the innermost western torus, the highly
redshifted HCN J=4--3 dense molecular emission is physically related
to the redshifted H$_{2}$O maser-emitting rotating disk.

\item
Comparison of J=4--3 and J=3--2 fluxes of HCN and HCO$^{+}$ emission
extracted with a 1.7 pc diameter circular aperture around the
HCN J=4--3 emission peaks in both the western and eastern tori
("W-peak" and "E-peak", respectively) suggests that dense
molecular line emission is optically thin in these small areas. 
High turbulence of molecular gas can reduce line opacity and reproduce
the observed molecular emission line properties there.
A physically counter-rotating dense molecular gas scenario in the NGC
1068 torus, driven by the collision of a compact gas clump with the
western part of the existing rotating torus from the opposite
direction, is a natural explanation if the interface of two
counter-rotating gas components is located close to these emission peaks.

\item
We applied the analytical model based on this assumption presented by
\citet{vol22} and found that the overall properties of our new HCN
J=4--3 and HCO$^{+}$ J=4--3 data are largely reproduced by this
model, as previously confirmed for other molecular lines, including
HCN J=3--2 and HCO$^{+}$ J=3--2.
However, 
the observed significant east-west asymmetry of dense molecular
line emission properties needs to be better reproduced with this model.
Furthermore, even using our highest spatial resolution ($\lesssim$1 pc)
molecular line data so far, (i) a predicted possible
$\sim$0.5--1.0 pc wide gap at the interface between two
counter-rotating gas components and (ii) high molecular gas excitation
caused by possible turbulence-induced shocks there are not
clearly detected. 
Even smaller beam-sized data in the future may help further
test the physically counter-rotating torus scenario in NGC 1068.

\item
Enhanced HCN abundance in the vicinity ($\lesssim$3--5 pc) of 
the luminous AGN in NGC 1068 is further supported by two additional 
observational facts:
(i) the blueshifted absorption wing toward the mass-accreting SMBH, 
likely originating from nuclear compact outflow activity, is stronger 
for HCN J=4--3 than for HCO$^{+}$ J=4--3, and
(ii) the HCN-derived torus dense molecular H$_{2}$ gas mass is 
a factor of $\sim$4--5 higher than the HCO$^{+}$-derived one, 
under the assumption of standard HCN and HCO$^{+}$ abundances 
for converting their luminosities to dense molecular H$_{2}$ gas mass.

\end{enumerate}

Our study demonstrated that the combination of
high-spatial-resolution, multiple J-transition molecular line data is
very powerful to scrutinize various properties of the compact dense
molecular torus in the nearby archetypal AGN NGC 1068.


\begin{acknowledgments}
We thank the anonymous referee for his/her valuable
comments, which helped improve the clarity of this manuscript.
This paper made use of the following ALMA data:
ADS/JAO.ALMA\#2022.1.00005.S and \#2018.1.00037.S.
ALMA is a partnership of ESO (representing its member states), NSF (USA) 
and NINS (Japan), together with NRC (Canada), NSC and ASIAA
(Taiwan), and KASI (Republic of Korea), in cooperation with the Republic
of Chile. The Joint ALMA Observatory is operated by ESO, AUI/NRAO, and
NAOJ. 
M.I. is supported by JP21K03632 and JP25K07359.
Y.H. is supported by JP25K07357.
Data analysis was in part carried out on the open use data analysis
computer system at the Astronomy Data Center, ADC, of the National
Astronomical Observatory of Japan. 
This research has made use of NASA's Astrophysics Data System and the
NASA/IPAC Extragalactic Database (NED) which is operated by the Jet
Propulsion Laboratory, California Institute of Technology, under
contract with the National Aeronautics and Space Administration. 
\end{acknowledgments}

%

\vspace{5mm}
\facilities{ALMA}
\software{CASA \citep{CASA22}}






\appendix

\section{Physical origin of the $\sim$400 km s$^{-1}$ redshifted
absorption feature seen in the HCO$^{+}$ J=4--3 spectrum}

An absorption feature, at $\sim$400 km s$^{-1}$ redshifted relative to
the systemic velocity (V$_{\rm sys}$ = 1130 km s$^{-1}$), is detected
in the spectra of HCO$^{+}$ J=4--3 (Figure \ref{fig:spectraPeakVopt}b
and \citet{gam25}), CO J=3--2 \citep{gam25}, and HCO$^{+}$ J=3--2
\citep{ima20}, toward the $\sim$260--350 GHz continuum emission peak
position. 
It is also seen in the position velocity diagrams of HCO$^{+}$ J=4--3,
HCO$^{+}$ J=3--2, and CO J=3--2, at the continuum emission peak
(Figure \ref{fig:PV}b,d and \citet{gam25}), but not seen for HCN
J=4--3 and HCN J=3--2 (Figure \ref{fig:PV}a,c).

The frequency of the $\sim$400 km s$^{-1}$ redshifted absorption
feature seen in the spectra of HCO$^{+}$ J=4--3 and J=3--2
agrees with that of HCN-VIB J=4--3 and J=3--2 lines, respectively,
because the rest frequency of the HCN-VIB line is 
$\sim$400 km s$^{-1}$ redshifted relative to that of HCO$^{+}$ 
(vibrational ground level, v=0), both at J=4--3 and J=3--2
(Figure \ref{fig:spectraPeak}a and \citet{ima20}).
If foreground absorbing dense molecular gas at the systemic velocity
(V$_{\rm sys}$ = 1130 km s$^{-1}$) is located close to the central
continuum-emitting AGN, HCN-VIB J=4--3 and J=3--2 lines can be
detected in absorption, given that in the close vicinity ($\lesssim$a
few pc) of the AGN in NGC 1068,
(1) HCN abundance is derived to be high
\citep[e.g.,][]{ima18a,ima20,vol22,but22}, and
(2) a sufficient amount of HCN can be vibrationally excited
(upper energy level E$_{\rm upper}$ = 1024 K) by AGN-origin infrared
radiation pumping \citep{sak10}.
In fact, the presence of a sufficient amount of vibrationally excited
HCN molecules (HCN-VIB) has been found in the vicinity of luminous
AGNs
\citep[e.g.,][]{sak10,ima13b,aal15a,aal15b,cos15,mart16,ima16b,ima16c,ima18b,fal19,fal21,sak21}.
The possible absorption dip by the HCN-VIB J=4--3 line is $\sim$0.5 mJy
beam$^{-1}$ (Figures \ref{fig:spectraPeak}a and
\ref{fig:spectraPeakVopt}b) or $\sim$7\% ($\tau$ $\sim$ 0.08) for the
$\sim$350 GHz continuum flux of $\sim$6.8 mJy beam$^{-1}$ (Table
\ref{tab:cont}).
A comparable level of absorption dip ($\tau$ $\sim$ 0.08) by the
HCN-VIB J=4--3 line has been detected toward the $\sim$350 GHz
continuum emitting mass-accreting SMBH in the luminous obscured AGN
NGC 1052 \citep{kam20}. 
The possible absorption dip by the HCN-VIB J=3--2 line toward the
$\sim$260 GHz continuum-emitting mass-accreting SMBH in NGC 1068 is
also $\sim$9\% ($\tau$ $\sim$ 0.09) \citep{ima20}.
This HCN-VIB absorption scenario can naturally explain the
non-detection of $\sim$400 km s$^{-1}$ redshifted absorption features
for HCN both at J=4--3 and J=3--2 (Figure \ref{fig:PV}a,c), because
there are no corresponding HCN-VIB lines there.
We thus argue that the absorption features centered at
V $\sim$ 1530 km s$^{-1}$ ($\Delta$V$_{\rm red}$ $\sim$ $+$400 km s$^{-1}$)
detected for HCO$^{+}$ J=4--3 and J=3--2, in the beam-sized spectra
toward the $\sim$260--350 GHz continuum emission peak (Figure
\ref{fig:spectraPeakVopt}b and \citet{ima20}) and position velocity
diagrams (Figure \ref{fig:PV}b,d), can be explained by HCN-VIB J=4--3
and J=3--2 lines at the systemic velocity, respectively.

The spatial extent of the $\sim$260--350 GHz continuum emission is
$\sim$1--2 pc (Figure \ref{fig:mom0} and \citet{gam25}) and the
Keplerian rotation velocity at $\sim$1 pc from the SMBH is $\sim$200
km s$^{-1}$ for the SMBH mass of M$_{\rm SMBH}$ $\sim$ 1 $\times$
10$^{7}$M$_{\odot}$, estimated for NGC 1068
\citep[e.g.,][]{gre96,hur02,lod03,gal23,gal24}.
Assuming that HCN molecules at $\sim$1 pc from the central AGN are
sufficiently vibrationally excited to v=1 level (HCN-VIB), then the
observed line width of $\sim$300--400 km s$^{-1}$ for the $\sim$400 km
s$^{-1}$ redshifted HCO$^{+}$ J=4--3 and J=3--2 absorption features
(Figure \ref{fig:spectraPeakVopt}c and \citet{ima20}) could be
explained by the HCN-VIB absorption scenario.

\citet{gam25} interpreted that the $\sim$400 km s$^{-1}$ redshifted
HCO$^{+}$ J=4--3 absorption feature originates in redshifted
($\sim$400 km s$^{-1}$) infalling gas in the close vicinity of the
central mass-accreting SMBH, because the $\sim$400 km s$^{-1}$ redshifted
absorption feature is seen also for CO J=3--2.
In this scenario, both the (i) detection of the absorption feature for
HCO$^{+}$ J=3--2 (Figure \ref{fig:PV}d) and (ii) non-detection for HCN
J=4--3 and HCN J=3--2 (Figure \ref{fig:PV}a,c) need to be explained.
At the same J transitions, the upper and lower energy levels are
comparable between HCN and HCO$^{+}$, but the critical density
(n$_{\rm crit}$) for HCN is a factor of $\sim$5 higher than that for
HCO$^{+}$ \citep{shi15}.
Thus, as long as we compare the same J=4--3 or J=3--2,
stronger absorption optical depth for HCO$^{+}$ than for
HCN could be explained by molecular gas whose density is high enough
to excite HCO$^{+}$ to J=3 and J=2, but not high enough to
sufficiently excite HCN to J=3 and J=2. 
However, it is not immediately obvious whether this scenario can
explain the observed significantly stronger absorption optical depth
for HCO$^{+}$ J=4--3 (lower energy level E$_{\rm lower}$ $\sim$ 26 K
and n$_{\rm crit}$ = 2--3 $\times$ 10$^{6}$ cm$^{-3}$ at 50--100 K)
than for HCN J=3--2 (E$_{\rm lower}$ $\sim$ 13 K and 
n$_{\rm crit}$ = 4--6 $\times$ 10$^{6}$ cm$^{-3}$) \citep{shi15}.

We ran RADEX calculations \citep{van07} for a foreground absorbing
gas cloud with a volume density of 10$^{5-6}$ cm$^{-3}$, kinetic
temperature of 50--150 K, and column density of 3 $\times$ 10$^{15}$
cm$^{-2}$ for HCO$^{+}$ \citep{gam25} and HCN, and line width of
300 km s$^{-1}$ (Figure \ref{fig:spectraPeakVopt}b).
If we adopt a background temperature of $\sim$3 K (i.e., the
cosmic microwave background temperature in the nearby universe), 
we find no solution to reproduce a factor of $\gtrsim$2 stronger
absorption optical depth for HCO$^{+}$ J=4--3 than for HCN J=3--2.

The continuum flux density at $\sim$350 GHz and $\sim$260 GHz is
6.8 mJy beam$^{-1}$ (for a synthesized beam of 0$\farcs$014 $\times$
0$\farcs$013) and 6.7 mJy beam$^{-1}$ (0$\farcs$019 $\times$
0$\farcs$017), respectively (Table \ref{tab:cont}).
We obtain a brightness temperature of $\sim$350 K from these continuum
data.
By inputting a background temperature of 350 K for our RADEX
calculations, we find that the absorption optical depth
for HCO$^{+}$ J=4--3 can be a factor of $\gtrsim$2 stronger than that
for HCN J=3--2, for a volume density of 10$^{5-6}$ cm$^{-3}$, kinetic
temperature of 50--150 K, and column density of 3 $\times$
10$^{15}$ cm$^{-2}$ both for HCO$^{+}$ and HCN, and a line width of
300 km s$^{-1}$.
For the kinetic temperature of 50--80 K, the absorption optical depths
of both HCO$^{+}$ J=4--3 and J=3--2 can be a factor of $\gtrsim$2
stronger than those of both HCN J=4--3 and J=3--2.
Thus, the observed detection of HCO$^{+}$ J=4--3 and J=3--2
absorption features, and non-detection of HCN J=4--3 and J=3--2
absorption features, can be quantitatively reproduced by an infalling
gas cloud with a $\sim$400 km s$^{-1}$ redshifted velocity \citep{gam25}.
 
The above comparison implicitly assumes that the HCN and HCO$^{+}$
abundances are comparable.
In the close vicinity of the AGN in NGC 1068, the HCN abundance is
derived to be significantly enhanced relative to HCO$^{+}$
\citep[e.g.,][]{ima18a,ima20,vol22,but22}, largely affected by the central AGN.
If a similarly enhanced HCN abundance is assumed for the infalling
gas cloud, then the stronger HCO$^{+}$ J=4--3 absorption compared to
HCN J=3--2 absorption becomes more difficult to explain.
However, it may be possible that the infalling gas cloud is not
HCN-abundance-enhanced, despite being located very close to the
central AGN, if the infalling cloud comes from a region sufficiently
far from the NGC 1068 nucleus and stays in the AGN vicinity only
for a short time period \citep{gam25}.
In this case, possible AGN effects on the abundance of the infalling gas
cloud are limited.
However, the elevated HCN abundance in NGC 1068 is reported not only
in the compact ($\lesssim$3--5 pc) torus region, but also in a more
extended ($\sim$100 pc) region \citep[e.g.,][]{ste94,nak18,tak19,but22,nak23}.
If the $\sim$400 km s$^{-1}$ redshifted HCO$^{+}$ J=4--3 and J=3--2
absorption features are due to an infalling HCO$^{+}$ gas clump, then
it is likely to come from a non-HCN-abundance-enhanced region, far
away ($\gtrsim$100 pc) from the central HCN-abundance-enhanced region
of NGC 1068.

In summary, both the HCN-VIB and infalling gas cloud scenarios could
reproduce the $\sim$400 km s$^{-1}$ redshifted absorption features
detected only for HCO$^{+}$ J=4--3 and J=3--2, but not for HCN J=4--3
and J=3--2.
The former scenario cannot immediately explain the $\sim$400 km s$^{-1}$
redshifted CO J=3--2 absorption feature \citep{gam25}, because there
is no HCN-VIB absorption line there.
It may come from a low-density ($\lesssim$10$^{4}$ cm$^{-3}$)
infalling gas cloud.
The latter scenario is feasible as long as the infalling gas cloud is
not HCN-abundance-enhanced.

If HCN-VIB lines are responsible for the $\sim$400 km s$^{-1}$
redshifted absorption features detected for HCO$^{+}$ J=4--3 and
J=3--2, then HCN-VIB lines from dense molecular gas, which is located in
the close vicinity of the central AGN in a direction perpendicular to
our line of sight and is strongly illuminated by the AGN radiation, are
expected to be detected in emission.
However, the detection of the HCN-VIB emission line is usually not
easy because it is swamped by the much brighter HCO$^{+}$ emission
line at the vibrational ground (v=0) level, which is separated only
by $\sim$400 km s$^{-1}$.
This is the case in the beam-sized and area-integrated spectra of the
W-torus (Figure \ref{fig:spectraPeak}b and \ref{fig:spectraArea}a).
The HCO$^{+}$ J=4--3 emission line profile is narrower in the E-torus
(Figure \ref{fig:spectraPeak}c and \ref{fig:spectraArea}b), but
the presence of the HCN-VIB J=4--3 emission line is not clear.

In Figure \ref{fig:spectraPeakVopt}c, an emission-like feature is
recognizable at $\sim$200--400 km s$^{-1}$ redshifted velocity
(V$_{\rm sys}$ + 200--400 km s$^{-1}$ or V = 1330--1530 km
s$^{-1}$), in the beam-sized HCN-VIB J=4--3 spectrum toward the
$\sim$350 GHz continuum emission peak (= position of the putative
mass-accreting SMBH).
This spectrum is likely to include emission from the innermost
($\lesssim$1 pc) western torus where redshifted 
$\sim$200--400 km s$^{-1}$ emission components are detected in the
intensity-weighted mean velocity (moment 1) maps of HCN J=4--3 (Figure
\ref{fig:mom1}a) and HCN J=3--2 (Figure \ref{fig:mom1}c and \citet{ima20}).
The redshifted emission-like feature in Figure
\ref{fig:spectraPeakVopt}c may originate from redshifted HCN-VIB
J=4--3 emission components at the innermost ($\lesssim$1 pc) western
torus.

\clearpage

\section{Gaussian fit of detected emission lines}

\begin{figure*}[!hbt]
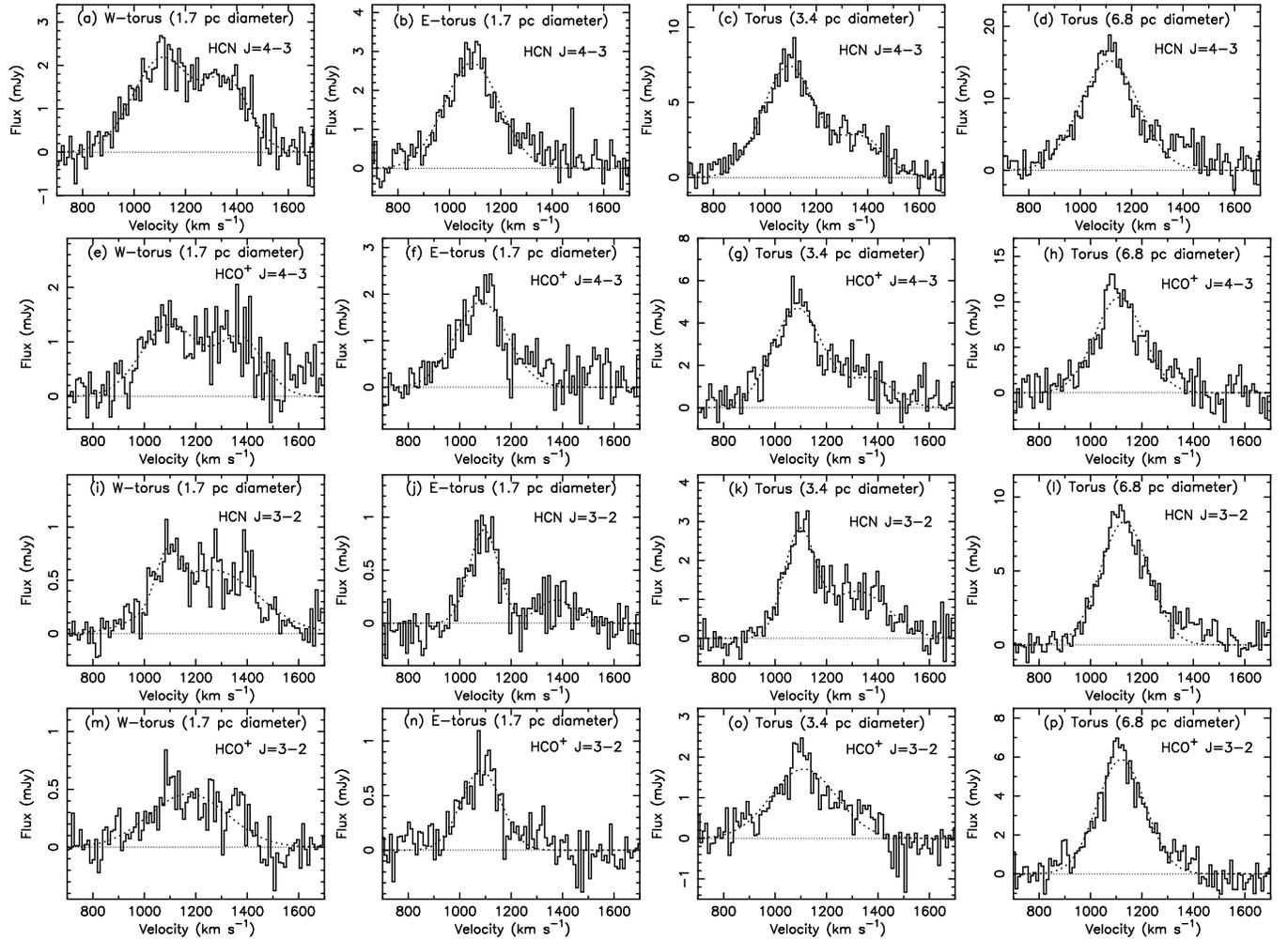

\begin{center}
\hspace{-0.4cm}
\includegraphics[scale=0.195,angle=-90]{fBa.eps} 
\includegraphics[scale=0.195,angle=-90]{fBb.eps} 
\includegraphics[scale=0.195,angle=-90]{fBc.eps}
\includegraphics[scale=0.195,angle=-90]{fBd.eps} \\
\includegraphics[scale=0.195,angle=-90]{fBe.eps} 
\includegraphics[scale=0.195,angle=-90]{fBf.eps} 
\includegraphics[scale=0.195,angle=-90]{fBg.eps}
\includegraphics[scale=0.195,angle=-90]{fBh.eps} \\
\includegraphics[scale=0.195,angle=-90]{fBi.eps} 
\includegraphics[scale=0.195,angle=-90]{fBj.eps} 
\includegraphics[scale=0.195,angle=-90]{fBk.eps} 
\includegraphics[scale=0.195,angle=-90]{fBl.eps} 
\includegraphics[scale=0.195,angle=-90]{fBm.eps} 
\includegraphics[scale=0.195,angle=-90]{fBn.eps} 
\includegraphics[scale=0.195,angle=-90]{fBo.eps} 
\includegraphics[scale=0.195,angle=-90]{fBp.eps} 
\end{center}
\vspace{+0.5cm}
\caption{
Gaussian fit (dashed line) of the detected HCN J=4--3 {\it (top)},
HCO$^{+}$ J=4--3 {\it (second top)}, HCN J=3--2 {\it (second bottom)},
and HCO$^{+}$ J=3--2 {\it (bottom)} emission lines at the W-torus 
(1.7 pc diameter) {\it (left)}, E-torus (1.7 pc diameter) 
{\it (second left)}, torus (3.4 pc diameter centered at the continuum
emission peak) {\it (second right)}, and the torus with an even wider
area (6.8 pc diameter) {\it (right)}.
The abscissa is the optical LSR velocity (in km s$^{-1}$), and the
ordinate is the flux density (in mJy).
The horizontal dotted straight line indicates the zero flux level.
Lower-frequency (redshifted) emission tails, relative to the molecular
line emission at the systemic velocity (V$_{\rm sys}$ = 1130 km
s$^{-1}$), are discernible in some of the spectra in
Figure \ref{fig:spectraArea}.
We apply two Gaussian fits to the HCN J=4--3, HCO$^{+}$ J=4--3, and HCN
J=3--2 emission lines at the W-torus and at the torus with a 3.4 pc
diameter aperture.
For the HCN J=3--2 emission line, a second redshifted emission
component is also added at the E-torus {\it (j)}.
For the remaining data, a single Gaussian fit is applied.
\label{fig:GaussFit}
}
\end{figure*}

Figure \ref{fig:GaussFit} displays the Gaussian fitting results for the
detected HCN J=4--3 and HCO$^{+}$ J=4--3 emission lines in the
area-integrated spectra at multiple torus positions, as well as
the HCN J=3--2 and HCO$^{+}$ J=3--2 emission lines detected in the
reanalyzed area-integrated spectra at the same positions.

\clearpage

\section{CS J=7--6 line}

\begin{figure*}[!hbt]
\begin{center}
\includegraphics[scale=0.5]{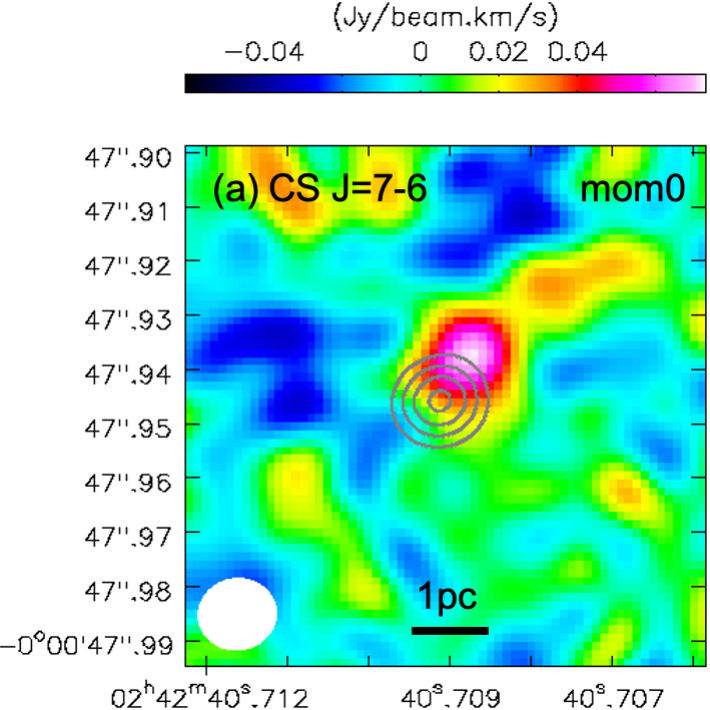} \\
\vspace{-0.7cm}
\includegraphics[scale=0.26,angle=-90]{fCb.eps} 
\includegraphics[scale=0.26,angle=-90]{fCc.eps} 
\includegraphics[scale=0.26,angle=-90]{fCd.eps} 
\end{center}
\vspace{+0.7cm}
\caption{
{\it (a)}: Integrated-intensity (moment 0) map of the CS J=7--6 line.
In the same way as in Figure \ref{fig:mom0}a, 
(i) the gray contours
indicate simultaneously obtained $\sim$350 GHz continuum emission with
50$\sigma$, 70$\sigma$, 90$\sigma$, and 110$\sigma$, 
(ii) the horizontal thick black bar corresponds to 1 pc at the distance
of NGC 1068, and 
(iii) the filled white circle indicates the synthesized beam size with 
0$\farcs$014 $\times$ 0$\farcs$013.
Observed area-integrated spectra around the CS J=7--6 line with
{\it (b)}: a 1.7 pc (24 mas) diameter circle centered at the W-peak,
{\it (c)}: a 1.7 pc diameter circle centered at the E-peak, and
{\it (d)}: a 3.4 pc (48 mas) diameter circle centered at the C-peak.
The abscissa is the observed frequency (in GHz) and the ordinate is the flux
density (in mJy).
Downward arrows are added at the expected frequency of the CS J=7--6
($\nu_{\rm rest}$ = 342.883 GHz) for the systemic velocity of
V$_{\rm sys}$ = 1130 km s$^{-1}$.
The horizontal dotted straight line indicates the zero flux level.
\label{fig:CS76}
}
\end{figure*}

Figure \ref{fig:CS76} presents an integrated-intensity (moment 0) map
and area-integrated spectra at multiple torus positions for the
CS J=7--6 ($\nu_{\rm rest}$ = 342.883 GHz) line.
The CS J=7--6 emission line is stronger (in the moment 0 map in
Figure \ref{fig:CS76}a) and broader (in the spectra in Figure
\ref{fig:CS76}b,c) in the western torus than in the eastern torus, as
also seen for HCN and HCO$^{+}$ at both J=4--3 and J=3--2 (Figures
\ref{fig:mom0} and \ref{fig:spectraArea}).

\end{document}